\documentclass[draft]{agujournal2019}
\usepackage{url} 
\usepackage{lineno}
\usepackage[inline]{trackchanges} 
\usepackage{soul}
\usepackage{amsmath}
\usepackage{booktabs}

\renewcommand{\vec}[1]{\mathbf{#1}}
\newcommand{\tens}[1]{\boldsymbol{\underline{\underline{#1}}}}

\usepackage{xcolor}

\usepackage{comment}

\draftfalse

\journalname{arXiv}

\begin{document}

\title{Coupling 3D geodynamics and dynamic earthquake rupture: fault geometry, rheology and stresses across timescales}

\authors{Anthony Jourdon\affil{1}, Jorge N. Hayek\affil{1}, Dave A. May\affil{2}, Alice-Agnes Gabriel\affil{2,1}}

\affiliation{1}{Department of Earth and Environmental Sciences, Ludwig-Maximillians-Universität München, Munich, Germany}
\affiliation{2}{Institute of Geophysics and Planetary Physics, Scripps Institution of Oceanography, UC San Diego, La Jolla, CA, USA}

\correspondingauthor{Anthony Jourdon}{anthony.jourdon@sorbonne-universite.fr}

\begin{keypoints}
\item We develop a new method to integrate 3D geodynamic models with dynamic rupture simulations.
\item Long-term rheology and evolution of fault geometry and pre-stress states crucially affect earthquake dynamics.
\item Geodynamically informed earthquake models highlight the need for detailed 3D fault modeling across time scales.
\end{keypoints}

\begin{abstract} 

Tectonic deformation crucially shapes the Earth's surface, with strain localization resulting in the formation of shear zones and faults that accommodate significant tectonic displacement. 
Earthquake dynamic rupture models, which provide valuable insights into earthquake mechanics and seismic ground motions, rely on initial conditions such as pre-stress states and fault geometry. 
However, these are often inadequately constrained due to observational limitations. 
To address these challenges, we develop a new method that loosely couples 3D geodynamic models to 3D dynamic rupture simulations, providing a mechanically consistent framework for earthquake analysis. 
Our approach does not prescribe fault geometry but derives it from the underlying lithospheric rheology and tectonic velocities using the medial axis transform.
We perform three long-term geodynamics models of a strike-slip geodynamic system, each involving different continental crust rheology. We link these with nine dynamic rupture models, in which we investigate the role of varying fracture energy and plastic strain energy dissipation in the dynamic rupture behavior.
These simulations suggest that for our fault, long-term rheology, and geodynamic system, a plausible critical linear slip weakening distance falls within $D_c \in [0.6,1.5]$.
Our results indicate that the long-term 3D stress field favors slip on fault segments better aligned with the regional plate motion and that minor variations in the long-term 3D stress field can strongly affect rupture dynamics, providing a physical mechanism for arresting earthquake propagation.
Our geodynamically informed earthquake models highlight the need for detailed 3D fault modeling across time scales for a comprehensive understanding of earthquake mechanics.
\end{abstract}

\section*{Plain Language Summary}
Tectonic deformation shapes the Earth's surface. 
Shear zones and faults represent the main structures accommodating movements between tectonic plates. 
Understanding earthquakes and mitigating seismic hazard heavily depends on accurate models of these faults and the stresses governing them. 
However, it is challenging to obtain precise details about fault geometry and stresses only from observations.
To overcome this issue, we developed a new method combining models of long-term geological evolution with earthquake simulations to provide a mechanically consistent way to obtain the stresses on and around faults and fault geometry to study earthquakes. 
This method relies on the mechanical properties of the Earth's crust and plates' tectonic motion over millions of years.
Using advanced 3D modeling software, we simulated the long-term evolution of strike-slip faults and extracted these faults to simulate dynamic rupture during an earthquake. 
We explored how different crustal mechanical behaviors and fault geometries influence earthquake dynamics.
Our study involved three long-term geological models and nine earthquake simulations. 
Our findings emphasize the importance of detailed 3D fault geometry and mechanical composition of the continental across time scales for a better understanding of earthquake mechanics, particularly in seismically active areas.

\section{Introduction}
Tectonic deformation plays a crucial role in shaping Earth's surface. 
Strain localization leads to the formation of shear zones at depth and faults at the surface, accommodating a significant portion of plate displacement within plate boundaries. 
Over millions of years, deformation can be considered as a spatially and temporally continuous process of visco-plastic strain localization \cite<e.g.,>[]{Gerya2019,Kirby1987,Ranalli1987,Ranalli1997}. 
At shorter timescales, strain localization involves the alternation of continuous visco-elastic deformation \cite<e.g.,>[]{Perfettini2004,Wahr1980} and discontinuous, almost instantaneous elasto-plastic deformation events rapidly releasing strain energy: earthquakes \cite<e.g.,>[]{Cocco2023,Gabriel2023faultsizedependentfractureenergy}. 
In active geodynamic systems, long-term tectonic forces and lithospheric responses pose the initial conditions governing earthquake nucleation, propagation, and arrest. 
However, long-term plate boundary formation and short-term earthquake mechanics are typically studied separately, and understanding their relationships across timescales and spatial scales remains a challenge \cite{lapusta2019modeling}.

The long-term visco-plastic mechanical behavior of the lithosphere heavily depends on rock rheology, which is influenced by chemical composition and the temperature field \cite<e.g.,>[]{Burov2011,Burgmann2008}, which are in turn impacted by the lithospheric geodynamic history \cite<e.g.,>[]{Beaumont2009,Jourdon2019,Jourdon2020,Manatschal2015}. 
The effect of continental crust rheology on strain localization has been extensively studied. It has been revealed that a lower continental crust deforming exclusively viscously (i.e., a weak crust) promotes diffuse deformation, low reliefs, and relatively low stress states. 
Conversely, continental crust with alternating layers of brittle/plastic and viscous/ductile behavior favors strain localization, supports high reliefs, and generates higher stresses \cite{Buck1991,Burov2011,Brun1999}. 
However, how the long-term rheology of continental crust influences earthquake mechanics remains unresolved.

Understanding earthquake dynamics is crucial for comprehending fault system interactions, assessing earthquake risks, and mitigating their impact. 
In tectonically active areas, the increasingly dense recording of seismic ground motion and geodetic deformation during and in between earthquakes contribute to establishing physical models to study earthquake dynamics \cite<, e.g.,>[]{Barbot2012,Jia2023}.
Among available approaches, dynamic rupture modeling provides forward models simulating earthquake evolution on fault surfaces non-linearly coupled to seismic wave propagation \cite<e.g.,>[]{Harris2018,Ramos2022}.
 However, this approach must rely on initial conditions, such as a mechanically self-consistent pre-stress state loading a fault before an earthquake and accurate 3D fault geometry. 
Constraining these initial conditions is a significant challenge \cite<e.g.,>[]{Tarnowski2022,Hayek2024}.
Nonetheless, the pre-stress state and the fault geometry significantly impact how earthquakes propagate  (e.g., crack- vs. pulse-like dynamics and subshear vs. supershear rupture speeds) and arrest \cite<e.g.,>[]{Kame2003,Bai2017} and the associated radiation of seismic waves and ground shaking \cite<e.g.,>[]{Harris2021,Taufiqurrahman2023}.

Recent studies have used long-term geodynamic models to constrain fault geometry and pre-stress states linked to earthquake dynamic rupture simulations \cite{Sobolev2017,VanZelst2019,Wirp2021,madden2021,vanZelst2022}. 
This approach, based on 2D visco-elasto-plastic long-term subduction simulations, embeds a long seismic cycle counting several thousands of years between two events, along with slip-rate dependent friction laws to generate stick-slip behavior on faults \cite{VanDinther2013,VanDinther2014,Herrendorfer2018}. 
Despite recent advances in modeling rupture dynamics in finite, deforming fault zones \cite{Benjemaa2007,Tavelli2020,Gabriel2021,Pranger2022,Hayek2023,Fei2023}, linking long-term geodynamic models to 3D dynamic rupture models typically requires constructing infinitesimally thin 2D fault surfaces from geodynamic volumetric shear zones.
Moreover, the rupture dynamics models revealed a strong dependency on lithological variations resolved by the long-term model, which are capable of slowing, stopping, or accelerating the rupture when passed and thus significantly altering co-seismic deformation.
However, limitations exist and include the extension from 2D to 3D to also resolve lateral, along-strike stress variations due to geometric and rheological variations \cite{Wirp2021}.

In this study, we first employ pTatin3D \cite{May2014,May2015}, a 3D long-term visco-plastic thermo-mechanical open-source finite element software, to simulate the geodynamic evolution of strike-slip deformation over geological timescales.
Using a new approach based on the medial axis transform, we extract 3D fault geometry, stress state, topography, and density as initial conditions for 3D dynamic rupture models performed with SeisSol, a short-term dynamic rupture and seismic wave propagation open-source discontinuous Galerkin software using unstructured tetrahedral meshes. This approach allows us to automatically extract 3D volumetric shear zones and map them into complex 2D fault interfaces.
We loosely couple (i.e., one-way link) volumetric fields such as stress and density from the long-term models and the dynamic rupture models and address the challenge of obtaining complex fault surfaces from a volumetric shear zone.

We investigate the relationships between long-term continental crust rheology and rupture dynamics by generating three long-term models with different crustal rheologies.
In nine geodynamically constrained 3D dynamic rupture models, we compare purely elastic media with varying fracture energy and models that additionally introduce off-fault deformation through plasticity.
We show that for our system, a dynamically plausible critical slip weakening distance falls within $D_c \in [0.6,1.5]$.
We establish a link between crustal rheology and rupture dynamics by comparing the rupture generated in a quartz-anorthite crust, a full quartz crust, and a full anorthite crust.
Produced earthquakes exhibit a shorter surface rupture length, a smaller rupture surface area, and less accumulated slip in a quartz-dominated crust than in an anorthite-dominated crust.
We also demonstrate how the long-term 3D stress field favors slip on fault segments better aligned with the regional plate motion and how minor variations in the long-term 3D stress field can strongly affect the rupture dynamics, providing a physical mechanism for arresting earthquake propagation.
\section{Long-term geodynamic modelling}
\label{sec:long-term}

\subsection{Governing equations}
\label{sub:lg-gov-eqs}
To simulate the long-term evolution of the deformation of the lithosphere, we utilize pTatin3D \cite{May2014,May2015}, a massively parallel visco-plastic finite element software.
This software solves for the conservation of momentum (Eq.~\eqref{eq:momentum}) and mass (Eq.~\eqref{eq:mass}) for an incompressible material:
\begin{align}
  \nabla \cdot (2 \eta(\vec u, p, T) \tens{\dot{\varepsilon}}(\vec u) ) - \nabla p + \rho(p,T) \vec{g} &= \vec 0,
  \label{eq:momentum} \\
  \nabla \cdot \vec u &= 0.
  \label{eq:mass}
\end{align}
Here, $\eta(\vec u, p, T)$ is the non-linear viscosity, while $\tens{\dot{\varepsilon}}(\vec u)$ is the strain rate tensor defined as:
\begin{equation}
  \tens{\dot{\varepsilon}}(\vec u) := \dfrac{1}{2} \left( \nabla \vec u + \nabla \vec u^T \right)
\label{eq:strain-rate}
\end{equation}  
with $p$ the pressure, $\rho$ the density, $\vec g$ the gravity acceleration vector and $\vec u$ the velocity.
The viscosity is highly dependent on temperature; thus, we solve the conservation of the thermal energy for an incompressible medium
\begin{equation}
  \rho_0 C_p \left( \dfrac{\partial T}{\partial t} + \vec u \cdot \nabla T \right) = \nabla \cdot (k \nabla T) + H_0 + H_s,
\label{eq:temperature}
\end{equation} 
with $T$ the temperature, $C_p$ the thermal heat capacity, $k$ the thermal conductivity and $\rho_0$ the reference density.
Heat sources considered here include an initial heat production $H_0$ depending on the lithology and simulating the radiogenic heat source of continental rocks and the heat dissipation due to the mechanical work
\begin{equation}
  H_s = \frac{2 \eta}{\rho_0 C_p} \tens{\dot{\varepsilon}}(\mathbf u) : \tens{\dot{\varepsilon}}(\mathbf u).
\end{equation}
In addition, to account for density variations due to temperature and pressure, we use the Boussinesq approximation and vary $\rho$ according to
\begin{equation}
  \rho(p,T) := \rho_0 (1 - \alpha (T - T_0) + \beta(p - p_0) ),
\label{eq:boussinesq}
\end{equation}
with $\rho_0$ the density of the material at $p=p_0$ and $T=T_0$, $\alpha$ the thermal expansion and $\beta$ the compressibility. 


\subsection{Rheological model} 
\label{sub:rheological_model}

The long-term rheology of the lithosphere is simulated using non-linear flow laws.
The ductile behavior is modelled using Arrhenius' type flow laws
\begin{equation}
  \eta_v (\vec u, p, T) := A^{-\frac{1}{n}} \left( \dot{\varepsilon}^{II}(\vec u) \right)^{\frac{1}{n} - 1} \exp{\left( \frac{Q + pV}{nRT} \right)},
\label{eq:eta-v}
\end{equation}
where $A$ the pre-exponential factor, $n$ the exponent and $Q$ the molar activation energy are material-specific parameters obtained from laboratory experiments, $R$ is the molar gas constant, $V$ the activation volume and
\begin{equation}
  \dot{\varepsilon}^{II}(\vec u) := \sqrt{ \dfrac{1}{2} \tens{\dot{\varepsilon}}(\vec u) : \tens{\dot{\varepsilon}} (\vec u) },
\label{eq:strain-rate-norm}
\end{equation}
the norm of the strain-rate tensor defined by Eq.~\eqref{eq:strain-rate}.

Moreover, the brittle behavior of the lithosphere is simulated using a Drucker-Prager yield criterion:
\begin{equation}
  \sigma_y (p) := C \cos{\phi} + p \sin{\phi}\,,
\label{eq:yield-stress}
\end{equation}
adapted to continuum mechanics by expressing it in terms of viscosity:
\begin{equation}
  \eta_p (\vec u,p) := \frac{\sigma_y (p)}{\dot{\varepsilon}^{II}(\vec u)},
\label{eq:eta-p}
\end{equation}
with $C$ the cohesion of the material and $\phi$ its friction angle.
In addition, we model the plastic softening with a linear decrease of the friction angle with the accumulation of plastic strain following
\begin{equation}
  \phi = \phi_0 - \frac{\epsilon_p - \epsilon_{\min}}{\epsilon_{\max} - \epsilon_{\min}} (\phi_0 - \phi_{\infty}),
\label{eq:friction-softening}
\end{equation}
with $\phi_0$ the friction angle of undamaged rocks, $\phi_{\infty}$ the friction angle of the fully softened rocks, $\epsilon_{\min}$ and $\epsilon_{\max}$ the amount of plastic strain between which the friction angle decreases and $\epsilon_p$ the cumulative plastic strain computed as
\begin{equation}
    \epsilon_p = \int \dot{\varepsilon}^{II}(\vec u) \, dt,
\label{eq:eps_p}
\end{equation}
when the material behaves plastically.

Finally, the viscosity of the lithosphere is evaluated with
\begin{equation}
  \eta(\vec u, p, T) = \min \big( \eta_v (\vec u, p, T), \eta_p(\vec u, p) \big).
\end{equation}


\subsection{Initial conditions} 
\label{sub:initial_conditions}

\begin{figure}[htb!]
\centerline{\includegraphics[width=0.99\textwidth]{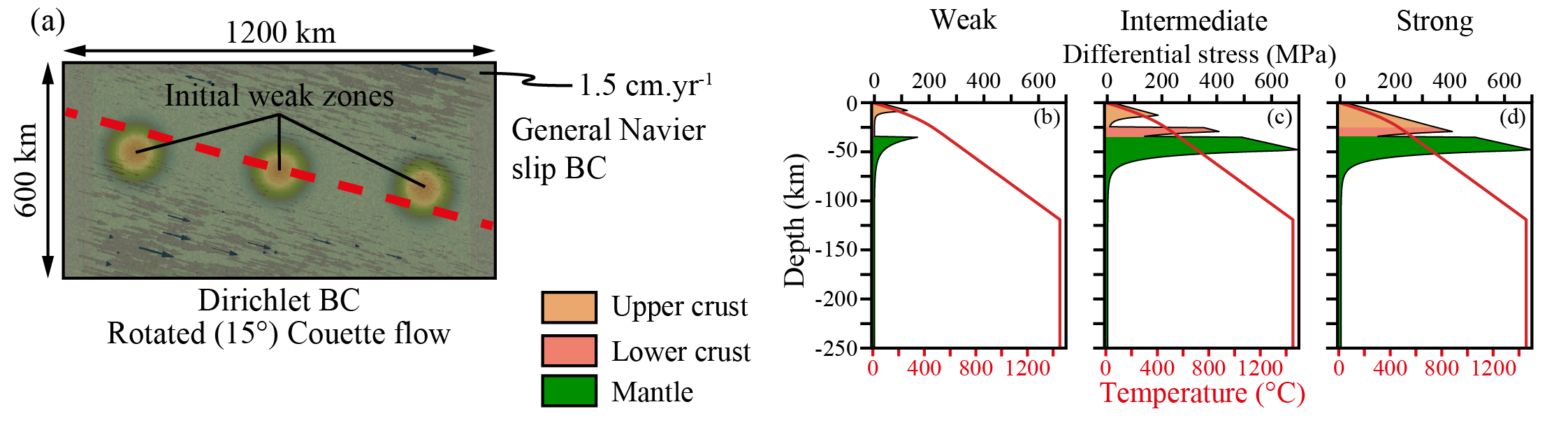}}
\caption{Initial and boundary conditions of the long-term geodynamic model.
(a) Map view of the domain. 
The colours show the initial weak zones imposed with a Gaussian repartitioning of random initial plastic strain that reduces the friction angle according to Eq.~\eqref{eq:friction-softening}.
The black arrows show the initial velocity field.
(b),(c),(d) Yield stress envelopes of the lithosphere computed for a strain-rate norm of $10^{-15}$ s$^{-1}$ for the weak (b), intermediate (c) and strong models (d).}
\label{fig:long-term-setup}
\end{figure}

The modelled domain $\Omega$ is represented in Cartesian coordinates, with $x$ and $z$ defining the horizontal plane, $y$ representing the vertical direction and $\vec{e_x}$, $\vec{e_y}$, $\vec{e_z}$ the three unit vectors defining the coordinate system. 
The origin of the domain is located at point $\vec O = [0, -200, 0]^T$, and its overall size is $1200 \times 200 \times 600$ km$^3$ (Figure \ref{fig:long-term-setup}a).
The model is divided into four initially flat layers, each representing specific geological materials and simulated using rheological properties reported in table~\ref{tab:geodyn-param}. 
The upper continental crust ranges from $0$ to $-25$~km, the lower continental crust from $-25$ to $-35$~km, the lithospheric mantle from $-35$ to $-120$~km, and the asthenosphere from $-120$~km.

The distinct rheological properties of each layer allow for the consideration of different behaviors and mechanical responses within the lithosphere and asthenosphere. 
By incorporating these rheological variations, the model aims to accurately capture the geodynamic processes occurring within the Earth's lithosphere.
In addition, to assess the importance of the lithosphere's long-term mechanical behavior on earthquake dynamics, three distinct lithosphere rheologies are considered.
The first model considers a weak continental crust (Figure \ref{fig:long-term-setup}b) entirely made of quartz \cite{Ranalli1987}.
The decoupling level between the crust and mantle is located around $-20$~km, and the whole lower crust exhibits a ductile behavior.
The second model considers a continental crust composed of two layers (Figure \ref{fig:long-term-setup}c), an upper crust made of quartz \cite{Ranalli1987} and a lower crust made of anorthite \cite{Rybacki2000}.
The anorthite in the lower crust introduces a stronger layer between the mantle and the upper crust, constraining the main decoupling level between the upper crust and the lower crust.
The third model considers a single-layer continental crust (Figure \ref{fig:long-term-setup}d) made of anorthite \cite{Rybacki2000}.
This model exhibits an almost fully plastic behavior from the surface to the mantle, removing any decoupling within the crust and between the crust and the mantle. 
In all models, the mantle is simulated with a dry olivine flow law \cite{Hirth2003}.

\begin{table}
  \begin{tabular}{l p{1.8cm} p{1.6cm} p{1.6cm} p{1.6cm} p{1.6cm} p{1.6cm} p{1.6cm} p{1.6cm} p{1.6cm}}
    \toprule
      & Units & \multicolumn{3}{c}{Upper crust} & \multicolumn{3}{c}{Lower crust} & Mantle\\
    \midrule
     & & M1 & M2 & M3 & M1 & M2 & M3 & & \\
    $A$ & MPa$^{-n}$.s$^{-1}$ 
    & $6.7 \times 10^{-6}$ & $6.7 \times 10^{-6}$ & $13.4637$ 
    & $6.7 \times 10^{-6}$ & $13.4637$ & $13.4637$ 
    & $2.5 \times 10^{4}$ \\ 
    $n$ & - 
    & $2.4$ & $2.4$ & $3$ 
    & $2.4$ & $3$ & $3$ 
    & $3.5$ \\
    $Q$ & kJ.mol$^{-1}$ 
    & $156$ & $156$ & $345$ 
    & $156$ & $345$ & $345$ 
    & $532$ \\
    $V$ & m$^3$.mol$^{-1}$ 
    & $0$ & $0$ & $3.8 \times 10^{-5}$ 
    & $0$ & $3.8 \times 10^{-5}$ & $3.8 \times 10^{-5}$
    & $8 \times 10^{-6}$ \\
    $C_0$ & MPa 
    & $20$ & $20$ & $20$
    & $20$ & $20$ & $20$
    & $20$ \\ 
    $C_{\infty}$ & MPa 
    & $5$ & $5$ & $5$ 
    & $5$ & $5$ & $5$
    & $5$\\
    $\epsilon_i$ & - 
    & $0$ & $0$ & $0$ 
    & $0$ & $0$ & $0$
    & $0$\\
    $\epsilon_e$ & - 
    & $0.5$ & $0.5$ & $0.5$
    & $0.5$ & $0.5$ & $0.5$
    & $0.5$\\
    $\beta$ & Pa$^{-1}$ 
    & $10^{-11}$ & $10^{-11}$ & $10^{-11}$
    & $10^{-11}$ & $10^{-11}$ & $10^{-11}$
    & $10^{-11}$\\
    $\alpha$ & K$^{-1}$ 
    & $3 \times 10^{-5}$ & $3 \times 10^{-5}$ & $3 \times 10^{-5}$ 
    & $3 \times 10^{-5}$ & $3 \times 10^{-5}$ & $3 \times 10^{-5}$ 
    & $3 \times 10^{-5}$\\
    $k$ & W.m$^{-1}$.K$^{-1}$ 
    & $2.7$ & $2.7$ & $2.7$
    & $2.85$ & $2.85$ & $2.85$ 
    & $3.3$ \\
    $H_0$ & $\mu$W.m$^{-3}$ 
    & $1.5$ & $1.5$ & $1.5$
    & $1.5$ & $0.3$ & $0.3$
    & $0$\\
    $\rho_0$ & kg.m$^{-3}$ 
    & $2700$ & $2700$ & $2700$
    & $2850$ & $2850$ & $2850$
    & $3300$\\
    \bottomrule
  \end{tabular}
\caption{Rheological and thermal parameters for the long-term geodynamic models. 
$A$, $n$, $Q$, and $V$ are the pre-exponential factor, the exponent, the activation energy, and the activation volume of the Arrhenius law, respectively (Eq. \ref{eq:eta-v}). 
$C_0$ and $C_{\infty}$ are the initial cohesion and the cohesion after softening.
$\epsilon_i$ and $\epsilon_e$ are the plastic strains at which softening starts and stops (Eq. \ref{eq:friction-softening}).
$\beta$ is the compressibility of the material and $\alpha$ is the thermal expansion coefficient for the Boussinesq approximation (Eq. \ref{eq:boussinesq}), $k$ the thermal conductivity, $H_0$ the initial radiogenic heat source and $\rho_0$ the initial density.}
\label{tab:geodyn-param}
\end{table}

The initial temperature field is the solution of the steady-state heat equation:
\begin{equation*}
  \nabla \cdot (k \nabla T) + H_0 = 0\,,
\end{equation*}
with the boundary conditions $T = 0^{\circ} \text{C } \forall y=0$ and $T=1450^{\circ} \text{C } \forall y=-200$ km.
In addition, to simulate an adiabatic thermal gradient maintained by mantle convection, we set the asthenospheric mantle conductivity to $70$ W.m$^{-1}$.K$^{-1}$ only for this initial steady-state solve.
Other parameters are reported in table \ref{tab:geodyn-param}.

\subsection{Boundary conditions} 
\label{sub:boundary_conditions}

In this study, we produce strike-slip deformation models by imposing far-field plate motion on the domain's vertical sides.
To avoid imposing a velocity discontinuity on the faces on which the velocity field changes polarity, we employ a newly developed method presented in \citeA{Jourdon2024}.
This method requires providing a direction in which the velocity must be constrained, and the stress tensor must be applied along faces.
In addition, this method can only be applied for velocity directions that are not orthogonal to the boundaries of the domain, explaining why we apply rotations of $\theta = 15^{\circ}$ in our boundary conditions.

Thus, to use this method, we divide the boundary of the domain into three sets: $\Gamma_D$ the set of boundaries using Dirichlet conditions, $\Gamma_N$ the set of boundaries using Neumann conditions, and $\Gamma_S$ the set of boundaries using Navier-slip conditions.
On faces of normal $\vec{e_z}$, we impose Dirichlet boundary conditions defined by a rotated horizontal Couette flow:
\begin{align*}
  \bar{u}_x &= \| \vec{\bar{u}} \| \left(\frac{2}{L_z} z - 1 \right)\,,\\
  \bar{u}_z &= 0\,,
\end{align*}
with $L_z$ the length of the domain in the $z$ direction and $\| \vec{\bar{u}} \|$ the relative velocity of plates.

On faces of normal $\vec{e_x}$ we impose the generalized Navier-slip boundary conditions defined by:
\begin{equation*}
  \vec u \cdot \vec{\hat{n}} = 0\,,
\end{equation*}
where $\vec{\hat{n}}$ is defined as the unit vector orthogonal to the velocity field imposed on the Dirichlet boundaries:
\begin{equation*}
  \vec{\hat{n}} =
  \begin{bmatrix}
    -\bar{u}_z\\
    0\\
    \bar{u}_x
  \end{bmatrix}
  \| \vec{\bar{u}} \|^{-1}.
\end{equation*}
In addition, we impose stress constraints in a coordinate system in which $\vec{\hat{n}}$ is one of the basis vectors.
To do so, let us denote
\begin{equation*}
  \tens{\Lambda} = 
  \begin{bmatrix}
    \vec{\hat{n}} & \vec{\hat{t}_1} & \vec{\hat{t}_2}
  \end{bmatrix}
  \,,
\end{equation*}
the matrix of the three orthogonal basis vectors forming a new coordinate system.
The imposed stress is thus defined as $\left(\tens{\Lambda} \tens{G} \tens{\Lambda}^T \right) \vec n$ where
\begin{equation*}
  \tens{G} := \tens{\mathcal{H}} \odot \left( \tens{\Lambda}^T \tens{\tau}_S \tens{\Lambda} \right)\,, 
\end{equation*}
with
\begin{equation*}
  \tens{\tau}_S := 2 \eta \tens{\dot{\varepsilon}}(\vec{\bar{u}}),
\end{equation*}
and $\tens{\mathcal{H}}$ a Boolean tensor designed to collect terms for which we apply a constraint ($\mathcal{H}_{ij} = 1$) and the terms that are treated as unknown ($\mathcal{H}_{ij} = 0$) which in our case is:
\begin{equation*}
  \tens{\mathcal{H}} = 
  \begin{bmatrix}
    0 & 1 & 1\\
    1 & 1 & 1\\
    1 & 1 & 0
  \end{bmatrix}
  .
\end{equation*}
More details about the method to apply such boundary conditions can be found in \citeA{Jourdon2024}.

In addition, we apply a Neumann free surface condition, $\tens{\sigma} \vec n = \vec 0$ to the top face and a constant value for $\vec u \cdot \vec n$ over the base of the domain ($\Gamma_{\text{base}}$) to ensure the compatibility constraint $\int_{\partial \Omega} \vec u \cdot \vec n \, dS = 0$ is satisfied.
The constant for the normal component of the velocity is referred to as a compensation velocity ($u_c$) and is computed as
\begin{align*}
    \int_{\partial \Omega} \vec u \cdot \vec n \, dS 
    &= \int_{\Gamma_\text{base}} \vec u \cdot \vec n \, dS + \int_{\partial \Omega  \backslash \Gamma_\text{base}} \vec u \cdot \vec n \, dS \\
    &= u_c \int_{\Gamma_\text{base}} 1 \, dS + \int_{\partial \Omega  \backslash \Gamma_\text{base}} \vec u \cdot \vec n \, dS \\
    &= 0.
\end{align*}


\section{Transforming volumetric shear zones into fault surfaces} 
\label{sec:shear-zones-2-faults}
The transformation of a volumetric shear zone into a fault surface poses a significant challenge to the Earth sciences community \cite<e.g.>[]{Duclaux2020,Neuharth2022,Pan2022} as well as industry \cite<e.g.>[]{Marfurt1998,Gersztenkorn1999,Gibson2005,Hale2012,An2021} for the generation of subsurface models.
Here, this poses a critical step in linking long-term geodynamic models with short-term earthquake dynamic rupture models.
Fundamentally, the question at hand is how to accurately capture the geometry of a complex 3D volume and represent it as a 2D surface. 

Varying approaches to surface reconstruction from cloud points have been proposed \cite{Berger2014SurfaceReconstruction}, with 3D problems being more challenging than 2D ones, especially with respect to scalability \cite{Tagliasacchi2016,Peters2018}. Conventionally, fault surface reconstruction is performed via manual fault interpretation. Automatic approaches involve the identification of discontinuities of seismic horizons through seismic attributes \cite{Marfurt1998,van2000seismic,BoeDaber2010,Song2012} or statistical approaches \cite{Wang2013}.
However, there are no established rules or definitive recipes for the required transformation process. Thus, we develop a method that involves condensing a relatively large volume, spanning a few kilometers, into a smaller set of points, typically within the range of a few tens or hundreds of meters. 
This condensed point set allows for the identification of a surface that can be further meshed using Delaunay triangulation.

To accomplish this transformation, we employ a geometric construct called the medial axis, which provides a framework for capturing the essential geometric features of the volumetric shear zone while reducing its representation to a simplified 2D surface.
The medial axis transform is sometimes referred to as ``skeletonization''.
By using the medial axis, which characterizes the central core or skeleton of the shear zone, we can effectively extract a subset of points that retain the essential characteristics of the original volume. 

In the subsequent steps, we employed a combination of Paraview \cite{Ahrens2005,Ayachit2015}, PyVista \cite{Sullivan2019}, and custom C code (for efficiency) to extract 2D fault surfaces from 3D shear zones.

\subsection{Shear zone identification} 
\label{sub:shear_zones_identification}

\begin{figure}[htb!]
\centering
\includegraphics[width=0.99\textwidth]{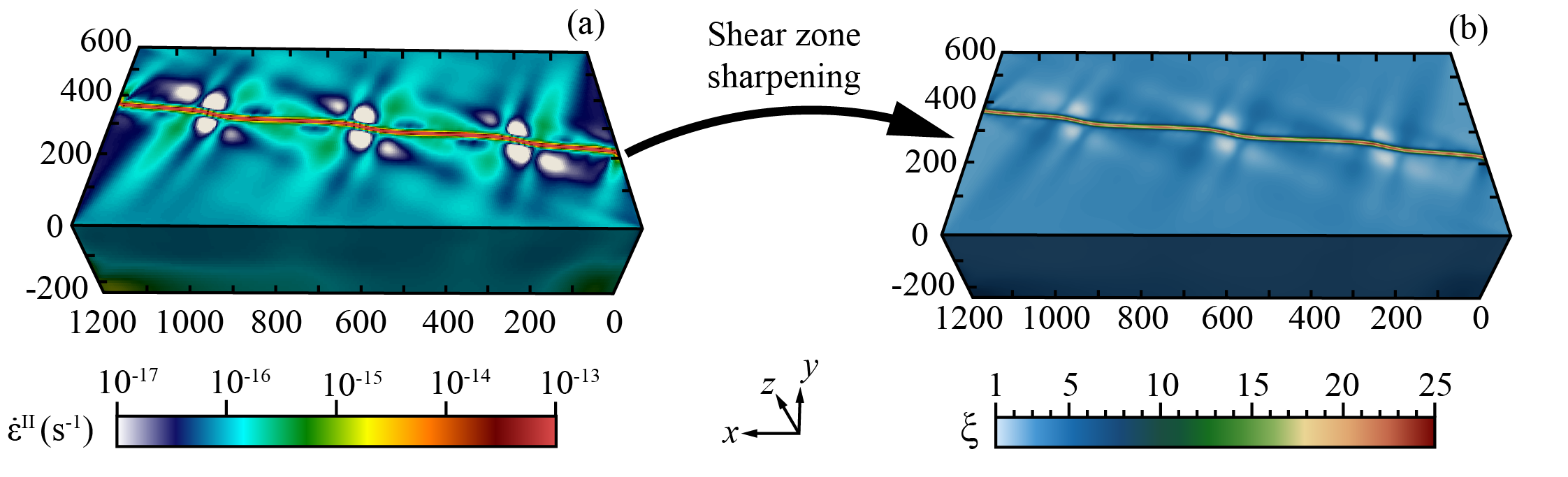}
\caption{3D long-term thermo-mechanical model. 
(a) Norm of the strain-rate tensor.
The red area shows the most localized active deformation, i.e., a shear zone.
(b) Result of applying the filter described in Eq~\eqref{eq:SZ-sharpening}.
}
\label{fig:SZ-sharpening}
\end{figure}

The initial step in extracting faults from shear zones involves determining the criteria for identifying what constitutes a shear zone. 
At its core, a shear zone can be defined as an area of localized strain, where the spatial derivative of displacement (for finite strain) or velocity (for active strain) is the most important. In this study, our focus is on active deformation, specifically on the strain rate.
Since velocity is a three-component vector, its spatial derivative corresponds to a nine-component tensor defined by Eq.~\eqref{eq:strain-rate}.
To assess the intensity of the strain rate tensor and identify regions of localized deformation, we employ the norm of the tensor described by Eq.~\eqref{eq:strain-rate-norm}.
This quantity represents the degree of localization of the deformation: higher values indicate more localized deformation. 
However, since the absolute value of this quantity is dependent on the velocities and distances within the domain, it is challenging to establish a universal threshold above which deformation can be considered localized. 
As a result, the strain-rate norm is used as a relative measure specific to each model, and its threshold may vary for different models. 
Experimental observations suggest that a localized shear zone can be established when there is a difference of approximately three to four orders of magnitude compared to areas with the lowest strain rates \cite<e.g.>[]{Brune2014,Liao2015,Sternai2016,LePourhiet2017,Neuharth2021,Jourdon2021}.
Nonetheless, we observe that our models with similar initial and boundary conditions tend to exhibit similar strain rate values.

To simplify the dimension reduction process, we apply an additional filter to the strain-rate norm, resulting in a new scalar field:
\begin{equation}
  \xi = 
    \exp{ 
      \left(
        \log_{10} \left(
          \dot{\varepsilon}^{II} 
        \right) 
        - \min \left( 
          \log_{10} \left( 
            \dot{\varepsilon}^{II} 
          \right) 
        \right) 
      \right) 
    }.
\label{eq:SZ-sharpening}
\end{equation}
The purpose of this fault indicator function is to enhance the visualization of shear zones and provide an initial treatment for the volume-to-surface transformation (Figure \ref{fig:SZ-sharpening}). 
By using the scalar field $\xi$, we construct surfaces of isovalues of $\xi$ that encapsulate the shear zones. 
For the model presented in Figure \ref{fig:SZ-MAT}a, a value of $\xi = 20$ was utilized.
However, for the same reasons that there is no universal value of $\dot{\varepsilon}^{II}$ to define a localized shear zone, there is no universal value of $\xi$, and a case-specific value must be chosen.
Additionally, we compute the outward-pointing normal vectors to these surfaces, which correspond to the shear zone boundaries (Figure \ref{fig:SZ-MAT}a).

By applying these techniques, we can effectively identify complex shear zones within the volumetric data and prepare them for further analysis and transformation into surface faults.

\subsection{Medial axis and surface meshing} 
\label{sub:medial_axis}

To reduce the dimensionality of the shear zones, we employ the shrinking-ball algorithm described in \citeA{Ma2012} to approximate the medial axis. 
However, shear zones extracted from numerical models are often characterized by surface roughness, which can introduce noise in the medial axis representation. 
Additionally, in regions such as the brittle-ductile transition within the lower continental crust or along the Moho, shear zones can flatten and spread over large distances, losing their relevance for earthquake dynamic rupture modeling and fault characterization as brittle objects.

\begin{figure}[htb!]
\centering
\includegraphics[width=0.99\textwidth]{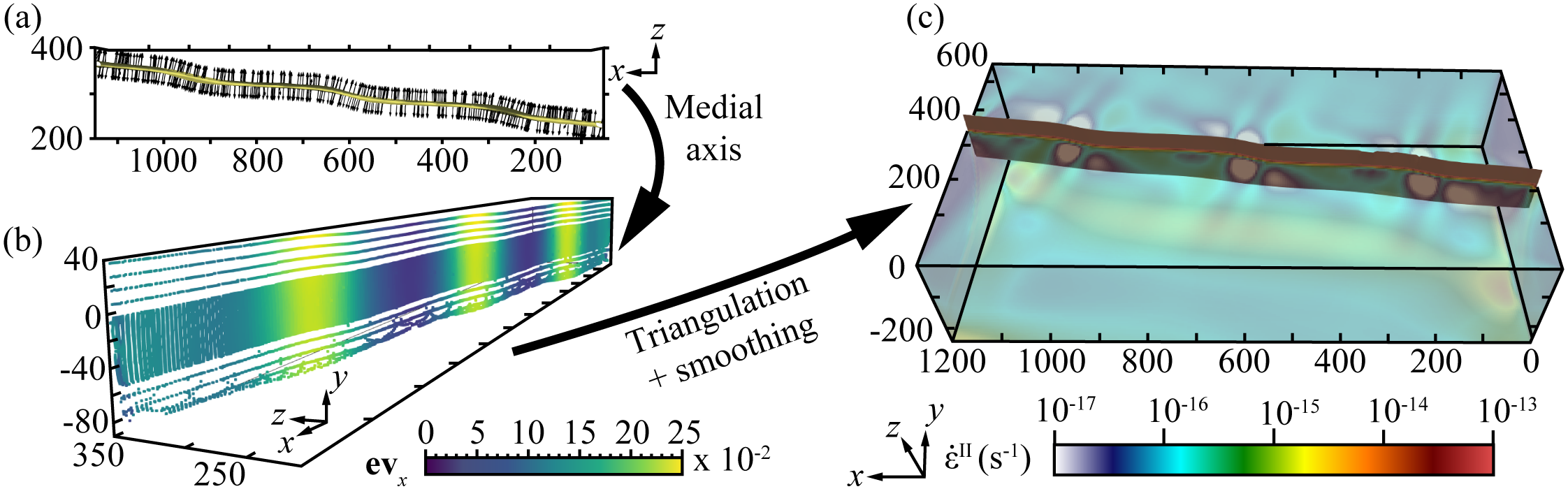}
\caption{(a) Contour of isovalue of $\xi = 20$ and normal vectors of this envelope at each point.
(b) Medial axis of the envelope shown in (a). 
Colours show the $x$ component of one of the eigenvectors of the covariance matrix computed with Eq.~\eqref{eq:covariance-matrix} illustrating the changes in orientation of the medial axis.
(c) Fault surface reconstructed from the medial axis using Delaunay triangulation and smoothing.
}
\label{fig:SZ-MAT}
\end{figure}

To address these issues and mitigate noise and effects associated with purely ductile deformation, we compute the geometric characteristics of the shear zone's medial axis. 
At each point with coordinates $\vec x = (x_1, x_2, x_3)^T$, we calculate the covariance matrix $\tens C$ of the spatial distribution of the set of points $\vec{X} = \{ \vec{x}_1, \vec{x}_2,..., \vec{x}_n \}$ within a sphere $\mathcal{S}$ of radius $r_s$ where
\begin{equation}
C_{ij} = \dfrac{1}{n} \sum_{k=1}^n (x_{ki} - \bar{x}_i)(x_{kj} - \bar{x}_j) \quad \forall \vec{x} \in \mathcal{S}(r_s),
\label{eq:covariance-matrix}
\end{equation}
and $\bar{\vec{x}}$ is the arithmetic mean
\begin{equation}
\bar{\vec{x}} = \dfrac{1}{n} \sum_{k=1}^n \vec{x}_k.
\label{eq:arithmetic-mean}
\end{equation}

The choice of $r_s$ is crucial as it determines the distance within which points are considered to contribute to the covariance matrix. 
However, to capture first-order orientation variations, the distance $r_s$ needs to be adjusted to represent a characteristic distance within which the orientation of the shear zone is representative of its surroundings.
After obtaining the covariance matrix for each point, we compute the eigenvectors associated with these matrices. 
The orientation of these eigenvectors provides information about the general orientation of the medial axis (Figure~\ref{fig:SZ-MAT}b), allowing us to remove points that deviate significantly from this orientation.

The remaining set of points is then utilized to create a surface using Delaunay triangulation. 
However, because the Delaunay triangulation attempts to connect all points with a given distance, this meshing process can result in a rough surface.
Therefore, to obtain a smooth 2D surface, we apply a Laplacian smoothing (Figure \ref{fig:SZ-MAT}c).

\subsection{Geometrically complex examples}
To demonstrate the applicability of the fault extraction method to different geodynamic scenarios and resulting fault types, here we consider a scenario involving a strike-slip shear zone splitting into two branches and an ocean-continent subduction producing two distinct shear-zones, a megathrust, and a conjugate thrust fault.

\begin{figure}[htb!]
\centering
\includegraphics[width=0.99\textwidth]{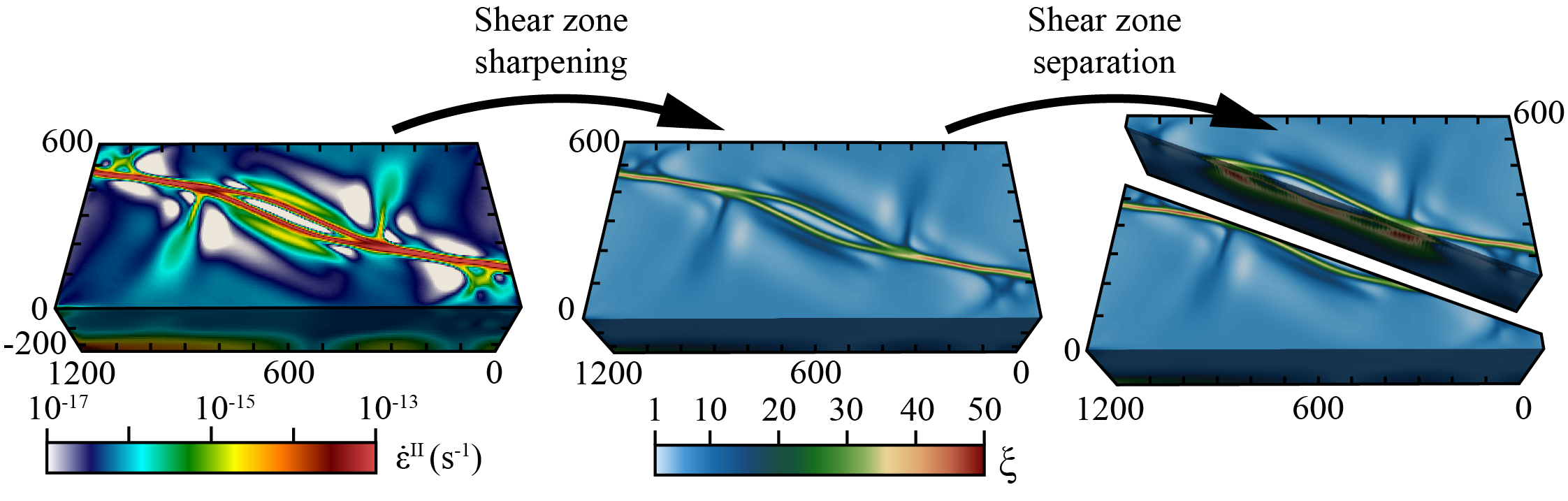}
\caption{
(a) Strain-rate norm  (Eq.~\ref{eq:strain-rate-norm}) showing the shear zone splitting.
(b) Shear zone sharpening using the $\xi$ filter (Eq.~\ref{eq:SZ-sharpening}).
(c) Volume splitting to isolate the two branches of the shear zone.
}
\label{fig:double-faults}
\end{figure}

\begin{figure}[htb!]
\centering
\includegraphics[width=0.99\textwidth]{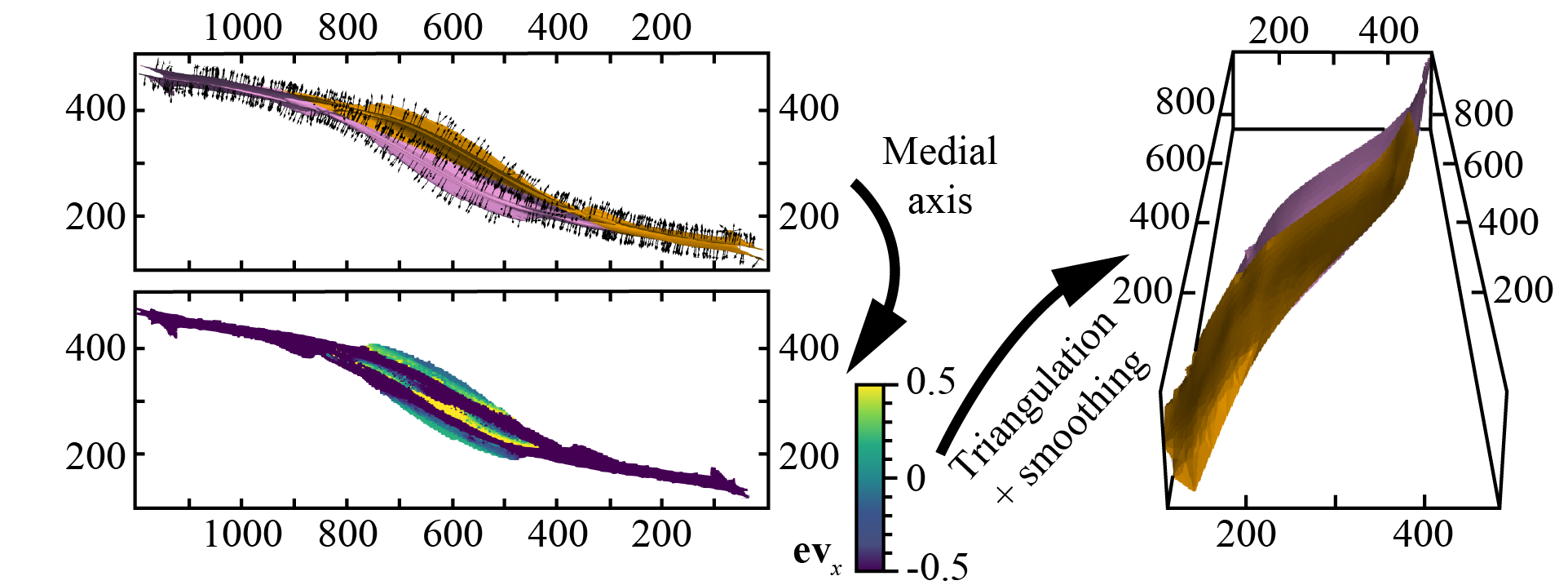}
\caption{
(a) Contour of the shear zone for $\xi = 20$.
The two colours indicate the contour from the two volumes represented in Figure~\ref{fig:double-faults}c.
The black arrows show the normal vectors to the surface of the contour $\xi = 20$.
(b) Medial axis of the two contours shown in panel (a).
The colours indicate the value of the $x$ component of one of the eigenvectors of the covariance matrix computed with Eq.~\eqref{eq:covariance-matrix} illustrating the changes in orientation of the medial axis.
(c) Fault surfaces of the two branches after Delaunay triangulation and Laplacian smoothing.
}
\label{fig:double-faults-2}
\end{figure}

In the case of a shear zone splitting into two branches (Figure~\ref{fig:double-faults}a), we first apply the $\xi$ filter to sharpen the shear zone (Figure~\ref{fig:double-faults}b) and split the volume into two pieces to isolate the two branches of the shear zone (Figure~\ref{fig:double-faults}c).
For each volume, we extract the contour of the shear zone using $\xi = 20$ (Figure~\ref{fig:double-faults-2}a) before computing the medial axis (Figure~\ref{fig:double-faults-2}b) and applying Delaunay triangulation and Laplacian smoothing (Figure~\ref{fig:double-faults-2}c).

\begin{figure}[htb!]
\centering
\includegraphics[width=0.99\textwidth]{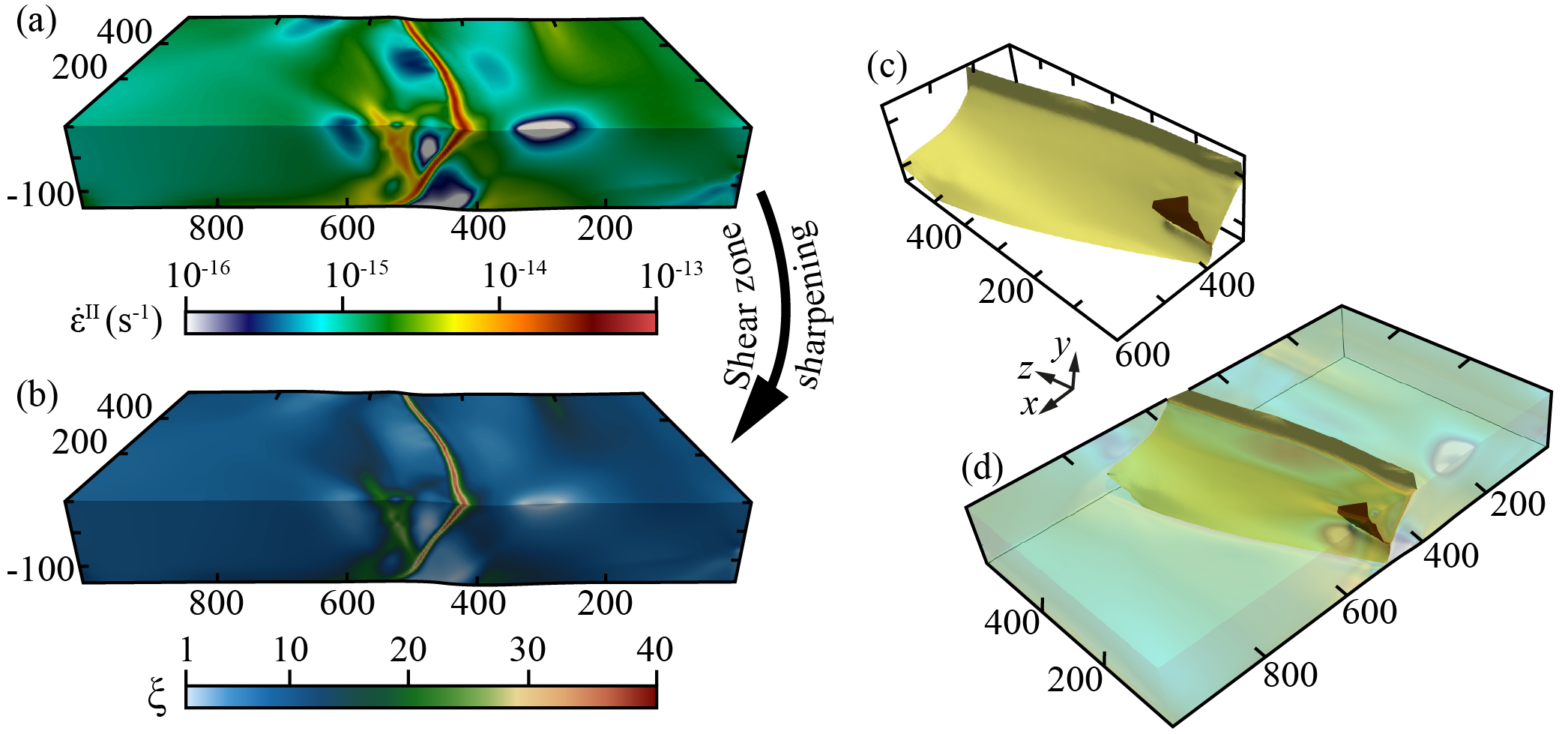}
\caption{Extraction of fault surface from a 3D subduction (collision) experiment including a megathrust and a conjugate thrust fault.
(a) Strain-rate norm (Eq. \ref{eq:strain-rate-norm}).
(b) $\xi$ value (Eq. \ref{eq:SZ-sharpening}).
(c) 3D view of the two extracted faults (red and yellow surfaces)
(d) 3D view of the two extracted faults within the domain. 
}
\label{fig:subduction-faults}
\end{figure}

In the case of the subduction model (Figure~\ref{fig:subduction-faults}a) we also apply the $\xi$ filter (Figure~\ref{fig:subduction-faults}b) to ease the extraction of the shear zone contour (Figure~\ref{fig:subduction-faults}c).

\section{Dynamic rupture modeling using 3D long-term geodynamic model data} 
\label{sec:dynamic_rupture_modeling}
\subsection{Governing equations} 
\label{sub:governing_equations}

To model 3D dynamic rupture and seismic wave propagation with high-order accuracy in space and time, we utilize SeisSol (\url{https://github.com/SeisSol/SeisSol})
\cite{Kaser2006,Pelties2012, Heinecke2014, Uphoff2017}, which employs fully non-uniform, unstructured tetrahedral meshes that statically adapt to geometrically complex 3D geological structures, such as non-planar mutually intersecting faults and topography. The code has been applied to model complex and/or poorly instrumented real earthquakes and earthquake scenarios in various tectonic contexts \cite<e.g.,>[]{Ulrich2019,Ramos2021,Biemiller2023,Tinti2021, Taufiqurrahman2023,Gabriel2023,Wang2024}.
SeisSol solves for the dynamic conservation of momentum 
\begin{equation}
  \rho \dfrac{\partial \vec u}{\partial t} - \nabla \cdot \tens{\sigma} = \vec 0
\label{eq:momentum-seisol}
\end{equation}
following the constitutive relationship
\begin{equation}
  \tens{\dot{\sigma}} - \lambda \nabla \cdot \vec u \tens{I} - G \tens{\dot{\varepsilon}} (\vec u) = \tens{0},
\label{eq:hook}
\end{equation}
with $\lambda$ and $G$ the Lamé parameters and $\rho$ the density of the material.
Eq.~\eqref{eq:momentum-seisol} is discretized using the discontinuous Galerkin method with arbitrary high-order derivative (ADER) time-stepping \cite{dumbser2006arbitrary}. SeisSol uses an end-to-end optimization for high-performance computing infrastructure \cite{Breuer2014,Heinecke2014,Uphoff2017,Krenz2021} and is verified in a wide range of community benchmarks \cite{Pelties2014} by the SCEC/USGS Dynamic Rupture Code Verification project \cite{Harris2018}.
The description of non-trivial initial conditions for SeisSol is provided by ASAGI (a pArallel Server for Adaptive GeoInformation, \citeA{Rettenberger2016}), an open-source library with a simple interface to access material and geographic datasets. ASAGI represents geoinformation on Cartesian meshes which are defined and populated with field files via a self-describing NetCDF file. ASAGI organizes Cartesian data sets for dynamically adaptive simulations by automatically migrating the corresponding data tiles across compute nodes as required for efficient access.


\subsection{Deviatoric stress and pressure} 
\label{sub:deviatoric_stress_and_pressure}

In addition to utilizing the long-term geodynamic model to obtain the fault geometry, we extract the 3D stress state to reconstruct self-consistent initial conditions for the dynamic rupture model.
The long-term geodynamic model solves for an incompressible visco-plastic Stokes flow, therefore the deviatoric stress tensor can be directly obtained from
\begin{equation}
  \tens{\tau}(\vec u, p) := 2 \eta(\vec u, p) \tens{\dot{\varepsilon}} (\vec u)\,,
\end{equation}
and already accounts for the long-term rheology (including the 3D temperature field), the geometry of the fault, the topography, and the volume forces.

In addition, although we obtain the pressure from the solution of Eqs.~\eqref{eq:momentum}~\&~\eqref{eq:mass}, this pressure satisfies the incompressibility constraint and thus can result in negative values. 
To avoid using negative values to represent the confining pressure and construct the full stress tensor, we utilize a different approach.
Based on \citeA{Jourdon2022} we compute the confining pressure $p_c$ related to the density structure in 3D described by
\begin{equation}
  \nabla \cdot (\nabla p_c) = \nabla \cdot (\rho \vec g),
\end{equation}
with the boundary conditions $p_c = 0$ at the surface and $\nabla p_c \cdot \vec n = \rho \vec g \cdot \vec n$ along the other boundaries, with $\vec n$ the outward pointing normal vector to the face.
Then, we compute the full stress tensor as
\begin{equation}
  \tens{\sigma} (\vec u, p, p_c) := \tens{\tau} (\vec u, p) -  p_c \tens{I}.
\end{equation}
To transfer the information from the long-term geodynamic model to the dynamic rupture model, we perform interpolation with iso-parametric $Q_1$ elements from the mesh of pTatin3D to a structured grid. This structured grid is used to interpolate values at the nodes of the unstructured tetrahedral mesh of the dynamic rupture model using ASAGI.

\subsection{Material parameters} 
\label{sub:material_parameters}
\begin{table}
  \centering
  \begin{tabular}{l p{1.8cm} p{1.6cm} p{1.6cm}}
    \toprule
    Parameters & Units &  \\
    \midrule
    $\lambda$ & Pa & PREM \\
    $G$ & Pa & PREM \\
    $\mu_s$ & - & 0.6\\
    $\mu_d$ & - & 0.1\\
    $D_c$ & m & $\{0.1,0.6,1,1.5,1.7\}$\\
    $C_d$ & MPa.km$^{-1}$ & 1\\
    $C_{\infty}$ & MPa & 1\\
    $y_r$ & km & 5\\
    $\phi_v$ & $^{\circ}$ & 30\\
    $C_v$ & MPa & 100 $\rightarrow$ 5\\
    \bottomrule
  \end{tabular}
\caption{3D dynamic rupture model parameters.
$\lambda$ and $G$ are the two Lam{\'e} parameters extracted from PREM \cite{Dziewonski1981}.
$\mu_s$ and $\mu_d$ are the static and dynamic friction coefficients, respectively.
$D_c$ is the critical slip distance.
$C_d$ is the on-fault cohesion slope, $C_{\infty}$ is the on-fault maximum cohesion, $y_r$ is the depth at which the maximum cohesion is reached (Eq. \ref{eq:on-fault-cohesion}).
$\phi_v$ is the volume friction angle for the models involving off-fault plasticity and $C_v$ the volume cohesion varying with the long-term plastic strain according to Eq.~\eqref{eq:friction-softening}. 
}
\label{tab:dr-param}
\end{table}

In dynamic earthquake rupture simulations, faults are typically idealized as infinitesimally thin interfaces separating distinct on- from off-fault rheologies { \cite<e.g.,>[]{Andrews2005,BenZion2005,TempletonRice2008,Dunham2011,Gabriel2013, Okubo2019,Hayek2023}. 

\subsubsection{On-fault parameters}\label{sec:methods-DR-onfault}

The static strength of crustal rocks can be high \cite{byerlee1978}. 
However, during co-seismic rupture, fault friction drops due to dynamic weakening processes \cite<e.g.,>[]{Kostrov1976,diToro2011,Kammer2024}.
We employ a linear slip-weakening friction law \cite{Andrews1976} describing the friction coefficient evolution with respect to the amount of slip along the fault
\begin{equation}
  \mu(S,D_c) := \mu_s - \frac{\mu_s - \mu_d}{D_c} \min(S,D_c),
\label{eq:slip-weakening}
\end{equation}
with $\mu_s$ the static friction coefficient, $\mu_d$ the dynamic friction coefficient, $D_c$ the critical slip distance and $S$ the slip defined as
\begin{equation*}
  S = \int_{0}^{t} \| \vec{u}(s) \| ds,
\end{equation*}
where $s$ is the fault surface.
We assume static and dynamic friction values from laboratory experiments \cite{byerlee1978,MooreLockner2007,diToro2011, scholz2019mechanics} in all our models.
Most of our models consider a uniform $D_c$. However, we also show two simulations with heterogeneous $D_c$ along the fault described by fractal hierarchical patches \cite{Ide2014}.

Moreover, to avoid a zero yield stress due to $p_c = 0$ at the surface we introduce an on-fault frictional cohesion $C (y)$ \cite{Harris2018} defined as
\begin{equation}
    C (y) := C_d (\max(y,y_r) - y_r) + C_{\infty},
\label{eq:on-fault-cohesion}
\end{equation}
with $C_d = 1$ MPa.km$^{-1}$ the slope, $y_r = 5$ km the depth at which the cohesion does not change anymore, $C_{\infty} = 1$ MPa the cohesion when $y < y_r$.
Note that $y_r < 0$ and $y$ decreases with depth.

Combining Eqs.~\eqref{eq:slip-weakening}~\&~\eqref{eq:on-fault-cohesion} gives the fault's yield strength
\begin{equation}
    \tau_f = - C(y) - \min(0,\sigma_n) \mu(S,D_c),
\end{equation}
allowing to evaluate if failure may occur.

In addition, to ensure consistency of the pressure and temperature-dependent stress tensor, we utilize the density extracted from the long-term geodynamic model.
Finally, utilizing the stress state extracted from the long-term geodynamic model and the dynamic rupture friction coefficients, the fault's relative strength $R$ can be evaluated from the ratio of the potential maximum stress drop to frictional strength drop \cite{Aochi2003} as
\begin{equation}
    R = \frac{|\tau_s| - \mu_d |\sigma_n|}{(\mu_s - \mu_d)|\sigma_n|},
\label{eq:fault-strength}
\end{equation}
where 
\begin{equation*}
  \sigma_n = \vec n \cdot \tens{\sigma} \vec n,
\end{equation*}
and 
\begin{equation*}
  \tau_s = \vec t \cdot \tens{\sigma} \vec n,
\end{equation*}
with $\vec n$ being the normal vector to the fault at a given point and $\vec t$ being the tangent vector in the direction of the slip.

\begin{figure}[htb!]
\centerline{\includegraphics[scale=1.0]{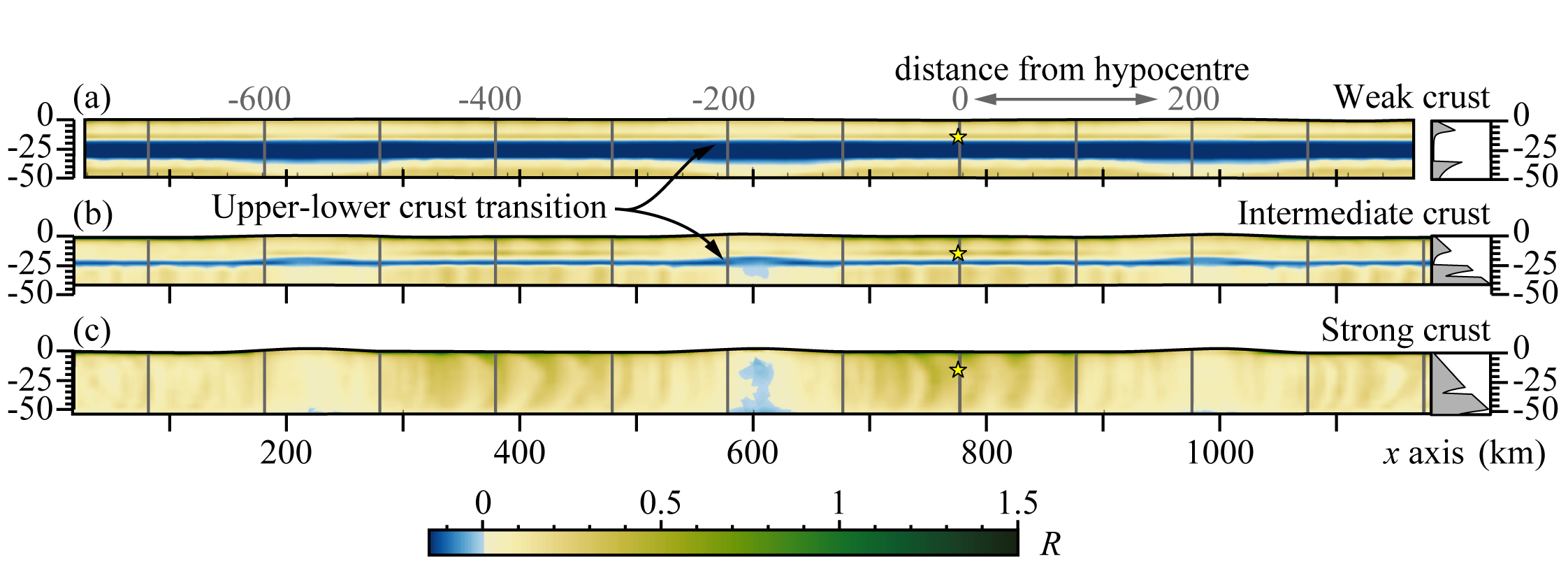}}
\caption{The initial $R$ ratio illustrating dynamic rupture relative fault strength computed with Eq.~\eqref{eq:fault-strength} for the long-term crustal rheology of a
(a) full quartz crust,
(b) quartz upper crust and anorthite lower crust,
(c) full anorthite crust.
The blue colours represent negative values of $R$ and indicate that $|\tau_s| < \mu_d |\sigma_n|$.}
\label{fig:strength-ini}
\end{figure}

Figure~\ref{fig:strength-ini} shows the initial $R$ value on the fault for each considered long-term rheology.
The faults and stress states extracted from the long-term geodynamic models with an upper crust composed of quartz (models 1 and 2) display an area of negative $R$ values corresponding to the location of the ductile decoupling layer arising directly from the long-term mechanical behavior of the crust.
The shallower part of the faults (i.e., above the surface of $R < 0$) can be interpreted as the seismogenic crust.
Conversely, the fault and stress state extracted from the model composed with a fully anorthite crust does not show  $R < 0$, indicating that the initial shear stress is everywhere above the dynamic strength of the fault, theoretically allowing rupture propagation on the entire fault surface.

\subsubsection{Off-fault parameters}
The dynamic rupture models that will be presented in sections \ref{subsubsec:ref-off-fault-damage}, \ref{subsec:weak-off-fault-damage}, and \ref{subsec:strong-off-fault-damage} include co-seismic off-fault plasticity \cite{Andrews2005,Wollherr2018}, allowing to capture volumetric plastic deformation around the fault.
As for the long-term geodynamic models, we also use a Drucker-Prager plasticity criterion described by Eq.~\eqref{eq:yield-stress} to define the yield stress.
However, for the dynamic rupture models, the friction angle in the volume is set to $\phi_v = 30^{\circ}$.
To ensure consistency between the long-term rheology, stress state,  finite deformation, and the dynamic rupture models parameters, the plastic cohesion for the off-fault plasticity is set using Eq.~\eqref{eq:friction-softening} in which we replaced $\phi$ by the volume cohesion $C_v$ varying between 100~MPa and 5~MPa.
The high value of 100~MPa ensures that the plastic yielding will not be reached far from the fault where no long-term deformation occurred.


\subsection{Nucleation} 
\label{sub:nucleation}

In dynamic rupture models, rupture nucleation requires only a small portion of the fault, a critical nucleation size \cite{RubinAmpuero2005}, to reach failure for rupture to initiate, even though other parts of the fault may remain below critical stress levels \cite{Kammer2024}.
Multiple techniques exist for nucleating dynamic earthquake ruptures, such as locally elevating shear stress, reducing the effective static frictional strength, or applying time-weakening forced nucleation strategies \cite{Andrews2004,Bizzarri2010,Hu2017,Harris2021}.
A spatially and temporally smooth nucleation patch avoids numerical artifacts \cite{Galis2015,Harris2018}.

Here we adopt a smooth time-weakening kinematic nucleation strategy by enforcing the time at which the friction coefficient reaches the dynamic value within a circular nucleation patch centered at the hypocenter. 
We choose the hypocenter to be located in the middle of one of the segments between the bends of our faults. 
We parameterize the forced rupture time within the nucleation patch of radius $r_c$ as 
\begin{equation*}
t_{forced}= \left(1-\exp{\left(\frac{r^2}{r^2-r_c^2}\right)}\right) t_{a} + t_{b}, 
\end{equation*}
where $r = || \vec x - \vec x_c ||$.
There, $r_c = 5$~km is the radius across which the friction coefficient smoothly decreases, $x_i$ the coordinates at the fault's surface and $x_{c_i}$ the coordinates of the hypocenter defined at $\vec{x}_c = [777, 275, -15.5]^T$ km, including scaling and offset from the start for the forced nucleation timing $t_{a}= 500$ and $t_{b}=0.2$.
Once the nucleation is sufficiently large, it is overtaken by spontaneous dynamic propagation of the rupture governed by the choice of friction law.


\section{Results} 
\label{sec:results}
To show how the long-term rheology, 3D stress-state, and fault geometry can influence the dynamics of the rupture during an earthquake, we first perform 3 geodynamic models with pTatin3d \cite{May2014,May2015} over 6 to 8 million years each with a different continental crust rheology and 14 dynamic rupture models with SeisSol \cite{Kaser2006,Pelties2012, Heinecke2014, Uphoff2017} varying rupture energy parameters with and without off-fault plasticity.
We briefly present six models in section \ref{subsubsec:Dc-variations} and, in more detail, three additional models with off-fault plasticity in sections \ref{subsubsec:ref-off-fault-damage}, \ref{subsec:weak-off-fault-damage} and \ref{subsec:strong-off-fault-damage}.

\subsection{Long-term geodynamic model}
\label{subsec:long-term-results}

\begin{figure}[htb!]
\centerline{\includegraphics[width=0.99\textwidth]{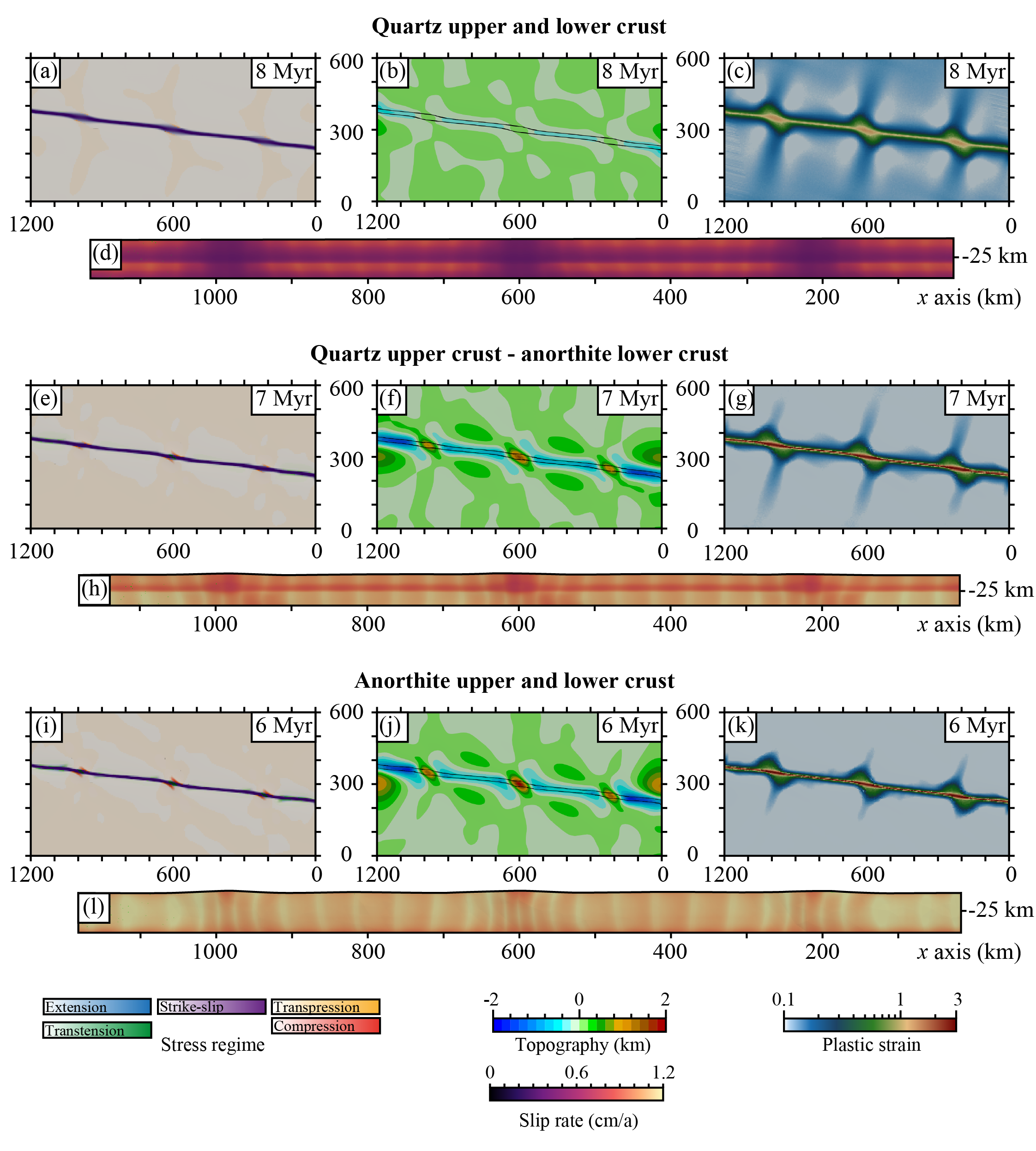}}
\caption{Long-term 3D geodynamic models.
(a-d) Single-layer model: quartz upper and lower crust.
(e-h) Two-layer model: quartz upper crust and anorthite lower crust.
(i-l) Single-layer model: anorthite upper and lower crust.
(a,e,i) Map view of the stress regime in active deformation zones.
(b,f,j) Map view of the topography.
(c,g,k) Map view of the plastic strain.
(d,h,l) Long-term slip rate computed on the fault's surface.}
\label{fig:long-term-maps}
\end{figure}

After 6 to 8 million years (Myr), the three long-term 3D geodynamic models develop a localized strike-slip shear zone (Figure \ref{fig:long-term-maps}a,e,i) with slight compression around the initial weak zone locations.
The long-term rheological properties of the crust significantly influence the topographic response to deformation (Figure \ref{fig:long-term-maps}b,f,j).
For instance, the model featuring a crust composed solely of quartz (the weaker model) exhibits topographic variations ranging from -0.5~km to 0.5~km, whereas the model with a crust composed entirely of anorthite (the stronger model) shows amplitudes ranging from -2~km to 2~km.
Furthermore, segments of the shear zone undergoing transpression/compression yield positive topography (mountain ranges), while segments experiencing transtension/extension result in negative topography (basins).

In the three models, the geodynamically modelled plastic strain (Figure \ref{fig:long-term-maps}c,g,k) illustrates the finite deformation, highlighting the advection and offset of the initial weak zones caused by the shear zone motion, as well as the accumulated strain, which delineates the highly localized shear zone at the center, alongside diffuse deformation oriented perpendicular to the main shear zone of the three models.
Those perpendicular diffuse deformation zones are inherited from the early phase of the model during which the strain starts to localize at the initial weak zones.
Moreover, although the geometry of the main shear zone remains relatively simple for all models, slight variations in orientation occur as it approaches the initial weak zone locations, leading to the formation of higher topography and a variation from strike-slip stress regime to transtension at the center of the shear zone and compression around.
These minor geometric variations contribute to form a non-planar shear zone and, thus, a non-planar reconstructed fault surface.

Once the corresponding fault surfaces are reconstructed, it becomes possible to evaluate the long-term slip rate across the faults (Figure \ref{fig:long-term-maps}d,h,l).
Once again, the crustal rheology significantly influences how much of the plate velocity imposed through boundary conditions is accommodated by the faults.
For instance, in the weakest model (quartz upper and lower crust), slip rates range from 0.4 cm.a$^{-1}$ to 0.7 cm.a$^{-1}$.
In contrast, the model with intermediate rheology (quartz upper crust and anorthite lower crust) accommodates slip rates from 0.7 cm.a$^{-1}$ to 1.2 cm.a$^{-1}$, while the strongest model (anorthite upper and lower crust) exhibits slip rates ranging between 0.8 cm.a$^{-1}$ and 1.2 cm.a$^{-1}$.
Additionally, in all three models, the long-term slip rate is influenced by geometric variations occurring at $x \approx {200,400,600}$~km, and where applicable, by the brittle-ductile transition between depths of -20 km and -25 km, characterized by lower slip rates.

\subsection{3D dynamic rupture: two-layered-crust models}
\label{subsec:quartz-anorthite-results}
 
\begin{figure}[htb!]
\centerline{\includegraphics[width=0.99\textwidth]{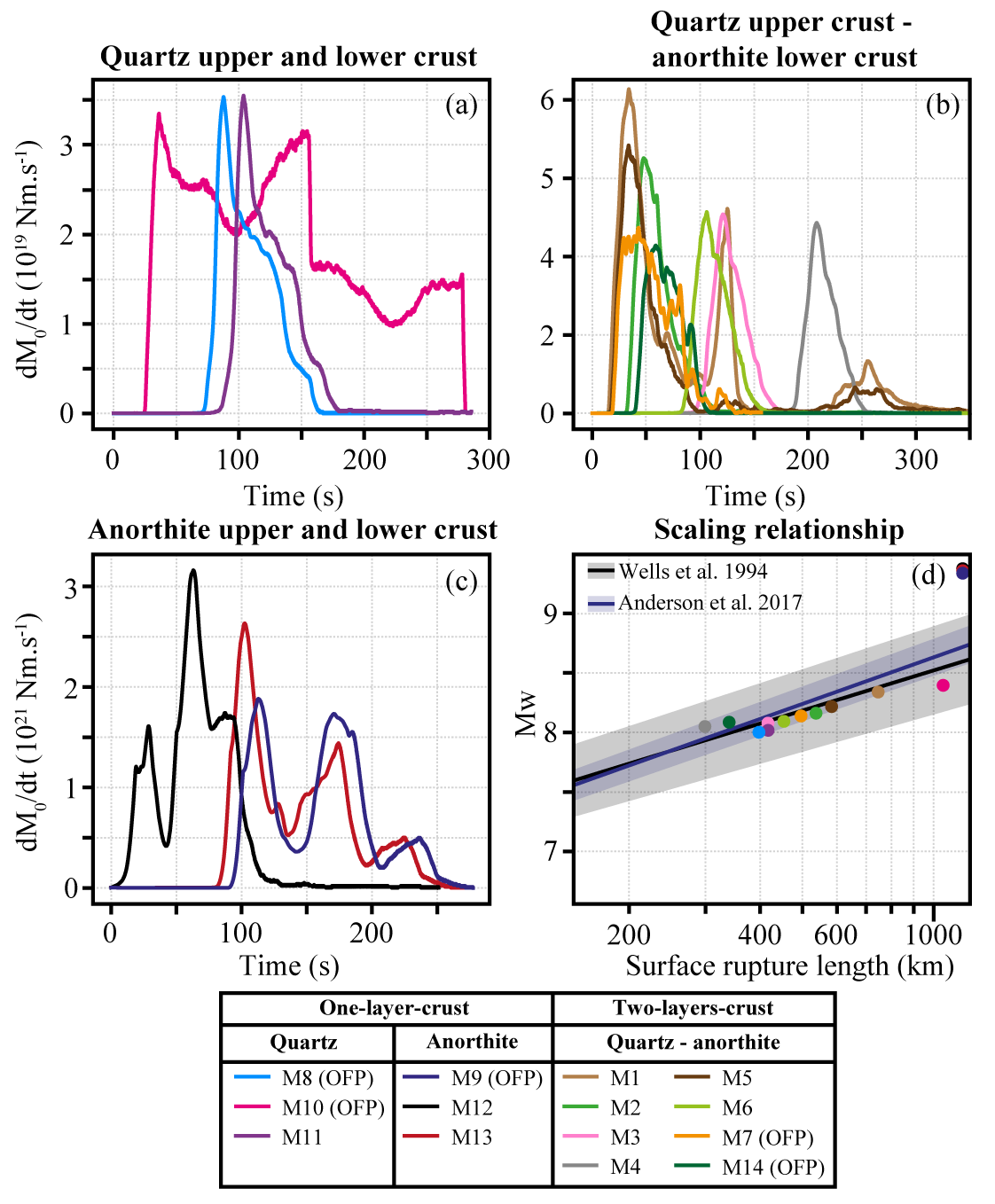}}
\caption{(a,b,c) Moment release rate $dM_0 / dt$ of the earthquake dynamic rupture simulations for 
(a) the single-layer quartz upper and lower crust models,
(b) the two-layer quartz upper crust and anorthite lower crust models,
(c) the single-layer anorthite upper and lower crust models.
(d) Scaling relationship between the surface rupture length and the magnitude $M_w$. The colored dots show our experiments and the black and blue lines show the empirical scaling relationships for strike-slip faults from \citeA{Wells1994} and \citeA{Anderson2017}, respectively.}
\label{fig:moment-scaling}
\end{figure}

\subsubsection{Coseismic fracture energy variation}
\label{subsubsec:Dc-variations}

\begin{figure}[htb!]
\centerline{\includegraphics[width=0.99\textwidth]{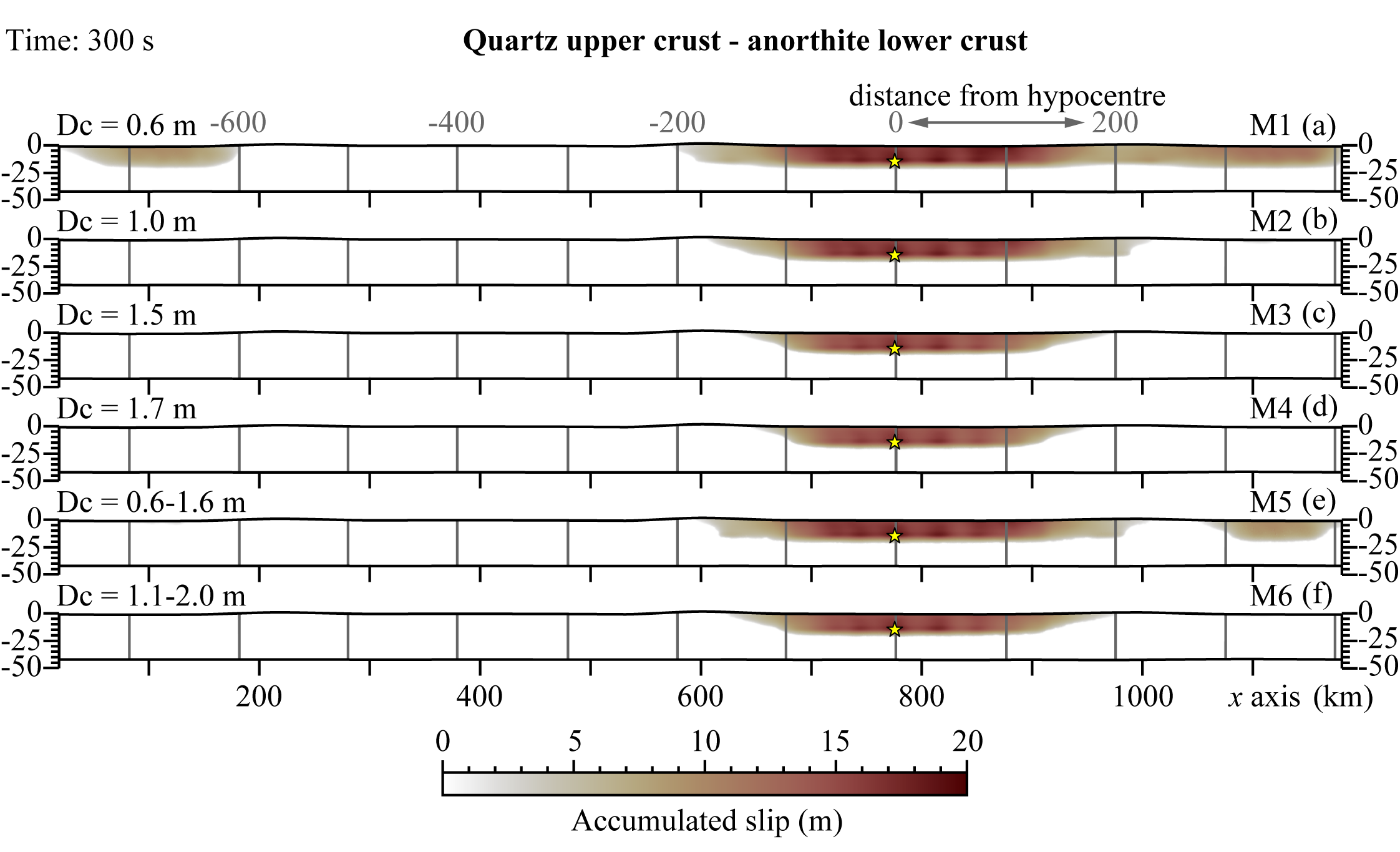}}
\caption{Fault slip for the simulated earthquakes with $D_c$ variations on the fault extracted from the two-layers-crust model with a quartz upper crust and anorthite lower crust. Elastic 3D dynamic rupture simulations without off-fault plasticity.
(a) M1: $D_c = 0.6$ m.
(b) M2: $D_c = 1$ m.
(c) M3: $D_c = 1.5$ m.
(d) M4: $D_c = 1.7$ m.
(e) M5: $D_c \in [0.6,1.6]$ m with fractal hierarchical patches.
(f) M6: $D_c \in [1.1,2]$ m with fractal hierarchical patches.}
\label{fig:reference-slip}
\end{figure}

\begin{figure}[htb!]
\centerline{\includegraphics[width=0.99\textwidth]{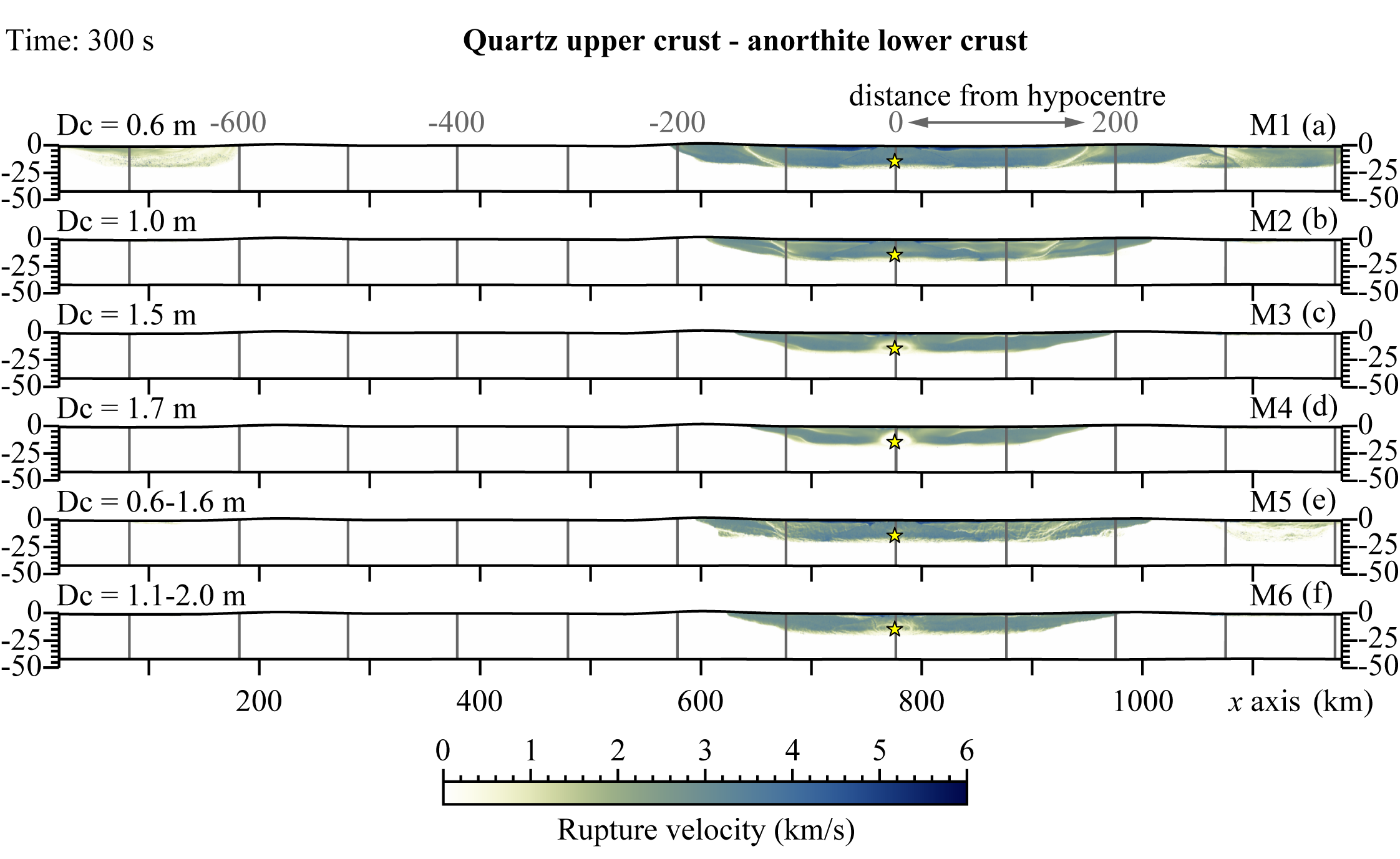}}
\caption{Rupture velocity for the simulated earthquakes with $D_c$ variations on the fault extracted from the two-layers-crust model with a quartz upper crust and anorthite lower crust. Elastic 3D dynamic rupture simulations without off-fault plasticity. 
(a) M1: $D_c = 0.6$ m.
(b) M2: $D_c = 1$ m.
(c) M3: $D_c = 1.5$ m.
(d) M4: $D_c = 1.7$ m.
(e) M5: $D_c \in [0.6,1.6]$ m in fractal hierarchical patches.
(f) M6: $D_c \in [1.1,2]$ m in fractal hierarchical patches.}
\label{fig:reference-velocity}
\end{figure}

During dynamic rupture, accumulating fault slip reduces the effective friction coefficient from its static to its dynamic value according to the linear slip-weakening friction law in Eq.~\eqref{eq:slip-weakening}. 
The critical slip-weakening distance ($D_c$) denotes the slip value over which friction decreases and correlates with the coseismic fracture energy dissipated during slip (Sec. \ref{sec:methods-DR-onfault}). 
Frictional parameters for dynamic rupture simulations are typically adopted from laboratory experiments. However, it is uncertain how valid it is to extrapolate results from the laboratory scale to the field scale \cite<e.g.,>[]{Cocco2023,Kammer2024}. 
Thus, we conducted six purely elastic models under a range of $D_c$ values.

Figure \ref{fig:reference-slip} illustrates the dynamic rupture fault slip that accumulated at the end of the simulation for each two-layer-crust model. 
All models show that dynamic rupture never penetrates the lower crust, which is consistent with the locally low on-fault strength parameter $R$ at depth (Figure \ref{fig:strength-ini}a). 

For model M1 with $D_c = 0.6$ m (Figure \ref{fig:reference-slip}a), along-strike rupture extends 200~km left and 400~km right of the hypocenter. 
A secondary 150~km long rupture segment, located 600~km left of the nucleation location, is dynamically triggered 200 seconds after the initiation of the rupture as shown in the moment rate release (Figure \ref{fig:moment-scaling}b).
Most of the slip accumulation occurs within a 200~km radius around the hypocenter, representing the segment between fault bends at $x \approx 600$~km and $x \approx 1000$~km. 
Figure \ref{fig:reference-velocity} displays the rupture velocity on the fault, showing a correlation between decreased accumulated slip and reduced rupture velocity at 100-150~km from the hypocenter, corresponding to locations where the fault orientation slightly changes and the long-term stress regime varies (Figure~\ref{fig:long-term-maps}). 
The M1 simulation results in an earthquake of magnitude $M_w = 8.34$ (Table~\ref{tab:dr-results}).

For model M2 with $D_c = 1$~m (Figure \ref{fig:reference-slip}b), dynamic rupture remains contained within the fault segment between fault bends at $x \approx 600$ km and $x \approx 1000$ km. 
The first 150 km around the hypocenter accumulate fault slip similar to the M1 ($D_c = 0.6$) m simulation.
The rupture velocity (Figure \ref{fig:reference-velocity}b)  slows near the fault's bends. 
The resulting earthquake magnitude is $M_w = 8.16$ (Table~\ref{tab:dr-results}).

Further increasing $D_c$ progressively reduces the rupture extent (Figure \ref{fig:reference-slip}c,d), while the accumulated slip near the nucleation area remains relatively constant (approximately $15$ to $20$ m) and moment rates show peak values that are very close to each other (Figure \ref{fig:moment-scaling}b). 
The moment magnitude of the modelled earthquakes decreases slightly ($M_w = 8.07$ for $D_c = 1.5$m and $M_w = 8.04$ for $D_c = 1.7$ m, see Table~\ref{tab:dr-results}). We observe delayed rupture initiation and slower propagation at the earthquake's onset (Figure \ref{fig:reference-velocity}c,d).

Simulations varying $D_c$ as fractal hierarchical patches \cite{IdeAochi2005} illustrate how non-uniform multi-scale critical slip-weakening distances may affect rupture dynamics.
Figure \ref{fig:reference-slip}e shows the accumulated slip for model M5 with $D_c$ varying from 0.6 m to 1.6 m, exhibiting characteristics of both $D_c = 1$ m (M2) and $D_c = 0.6$ m (M1) simulations and resulting in a magnitude $M_w = 8.21$ (Table~\ref{tab:dr-results}). 
However, varying $D_c$ introduces rheological variations on the fault's surface, observable at rupture area edges. 
Variations in critical slip-weakening distance influence the rupture velocity (Figure \ref{fig:reference-velocity}e), adding smaller scale perturbations during the propagation of the rupture compared to a uniform $D_c$. 

Finally, the simulation M6 with $D_c$ varying from 1.1 m to 2 m (Figure \ref{fig:reference-slip}f) shows slip patterns similar to the M3 simulation ($D_c = 1.5$ m) but with a delayed rupture initiation by 90 seconds. 
In this scenario, the earthquake magnitude is $M_w = 8.09$ (Table~\ref{tab:dr-results}).

These simulations suggest that for our fault, long-term rheology, and geodynamic system, a dynamically plausible critical slip weakening distance falls within $D_c \in [0.6,1.5]$. 
Lower values may yield unrealistically large earthquakes, while higher values delay the rupture initiation, indicating insufficient stress from the long-term geodynamic evolution to initiate rupture. 
Moreover, variations in uniform $D_c$ between 0.6~m and 2~m do not affect the first-order behavior of dynamic rupture and are thus omitted from the next experiments.

\subsubsection{Off-fault plasticity}
\label{subsubsec:ref-off-fault-damage}
\begin{figure}[htb!]
\centerline{\includegraphics[width=0.99\textwidth]{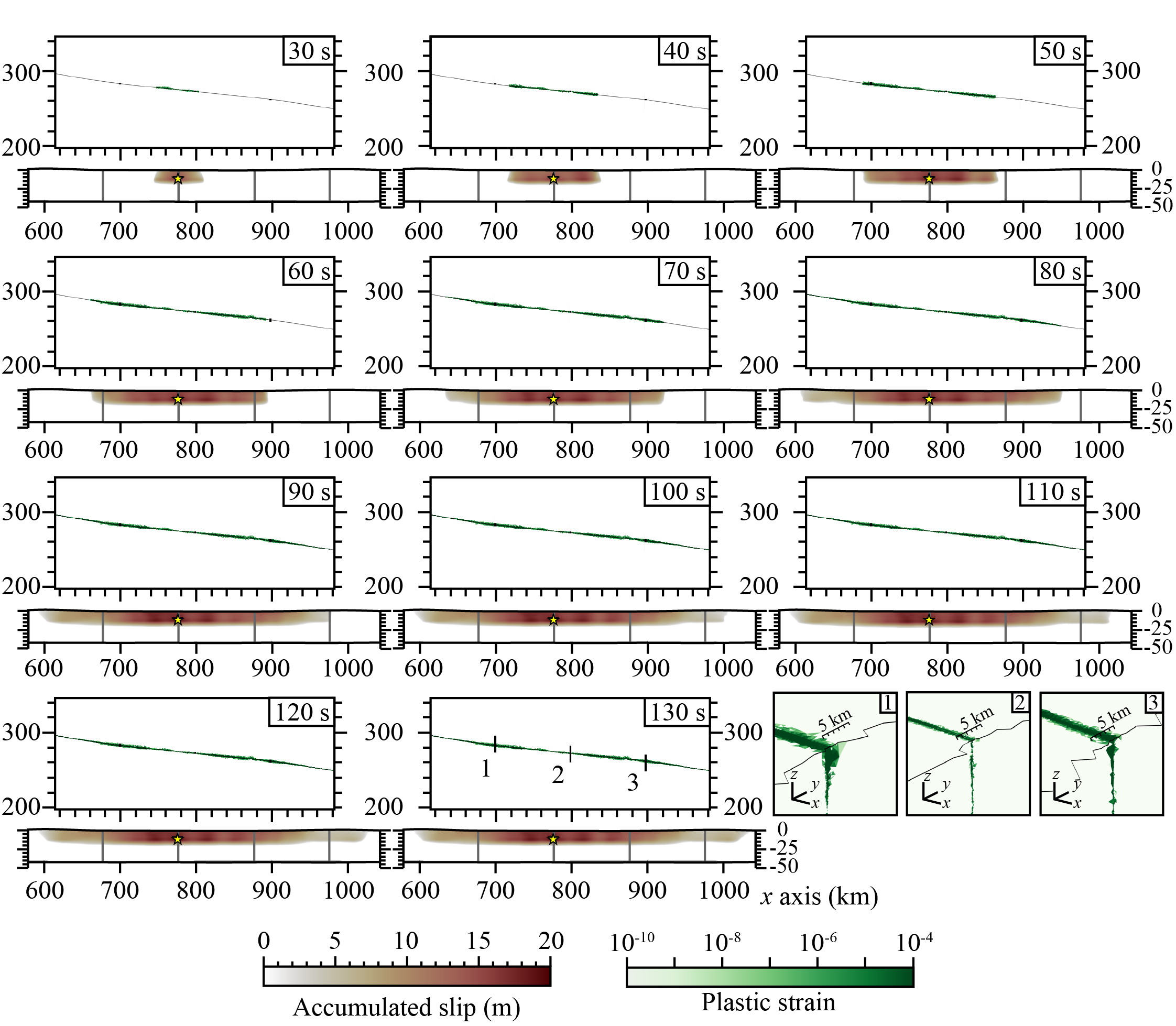}}
\caption{Evolution of the on-fault accumulated slip and off-fault plastic strain for $D_c = 0.6$ m in the model with a quartz upper crust and an anorthite lower crust.
The trace of the fault (thin black line) is represented in map view with the accumulated plastic strain.
The three zoomed cross-sections (1, 2, 3) are indicated on the map.
Below the map view, a zoom of the fault's surface between $x \approx 600$ km and $x \approx 1000$ km shows the accumulated slip at the given time.
The vertical grey lines are spaced every 100 km starting from the hypocenter.
The length scale between the map and the fault's surface are the same.}
\label{fig:reference-off-plast}
\end{figure}

We next introduce off-fault plasticity into our dynamic rupture models. This requires loosely coupling stresses everywhere in the 3D domain, not restricted to the fault surfaces. 
Based on calibration performed on the critical slip-weakening distance, we conducted a simulation with $D_c = 0.6$ m, with an upper crust composed of quartz and a lower crust composed of anorthite.
Figure \ref{fig:reference-off-plast} illustrates the evolution of the rupture on the fault through accumulated slip and in the volume through off-fault plastic deformation. 
After 30 seconds of simulation, the rupture spans a radius of 40~km around the hypocenter, with plastic deformation accumulating on the surface within a roughly 1~km thick zone and around the fault plane at depth within a thinner, approximately 500~m thick zone. 
The rupture propagates symmetrically on both sides of the hypocenter, and plastic deformation localizes in a broader area around the fault. It exhibits a positive flower-like structure with a thin (approximately 500~m) plastic deformation zone at depth and a thicker (around 2 to 3~km) deformation zone at the surface.

After 70 seconds, the rupture reaches the fault's bends located around the initial long-term weak zones ($x \approx {600,1000}$ km), breaking the symmetry of the propagation. 
Consequently, the rupture accumulates less slip and plastic strain, gradually fading and ultimately stopping after 130 seconds.
The moment rate decreases drastically at $\sim 80$ seconds, showing that the mechanical work required to pass through the fault's bend increases and consumes energy (Figure \ref{fig:moment-scaling}b). 
The slip pattern compares well to the elastic simulation with $D_c = 1$~m. 
The resulting magnitude is $M_w = 8.19$ for a surface rupture length of 550 km and a rupture area of 9120 km$^2$ (Table \ref{tab:dr-results}).

Compared with the model M1 (Figure~\ref{fig:reference-slip}a) using the same $D_c$ parameter but without off-fault plasticity, this model shows that all the characteristics of the rupture have been reduced (Table~\ref{tab:dr-results}, M7).
This behaviour is due to the plastic strain dissipating some of the earthquake energy, thus reducing the energy left for the rupture itself.

\subsection{Dynamic rupture: single-layer crust models}
To study the relationships between the long-term rheology and the dynamics of the rupture, we performed dynamic rupture experiments with and without off-fault plasticity.
However, in the following section, we only discuss the simulation with off-fault plasticity for the quartz upper and lower crust model, which is assimilated as a weak crust in terms of long-term rheology, and for the anorthite upper and lower crust model, which is assimilated as a strong crust. 

\subsubsection{Quartz upper and lower crust (weak rheology)}
\label{subsec:weak-off-fault-damage}
\begin{figure}[htb!]
\centerline{\includegraphics[width=0.99\textwidth]{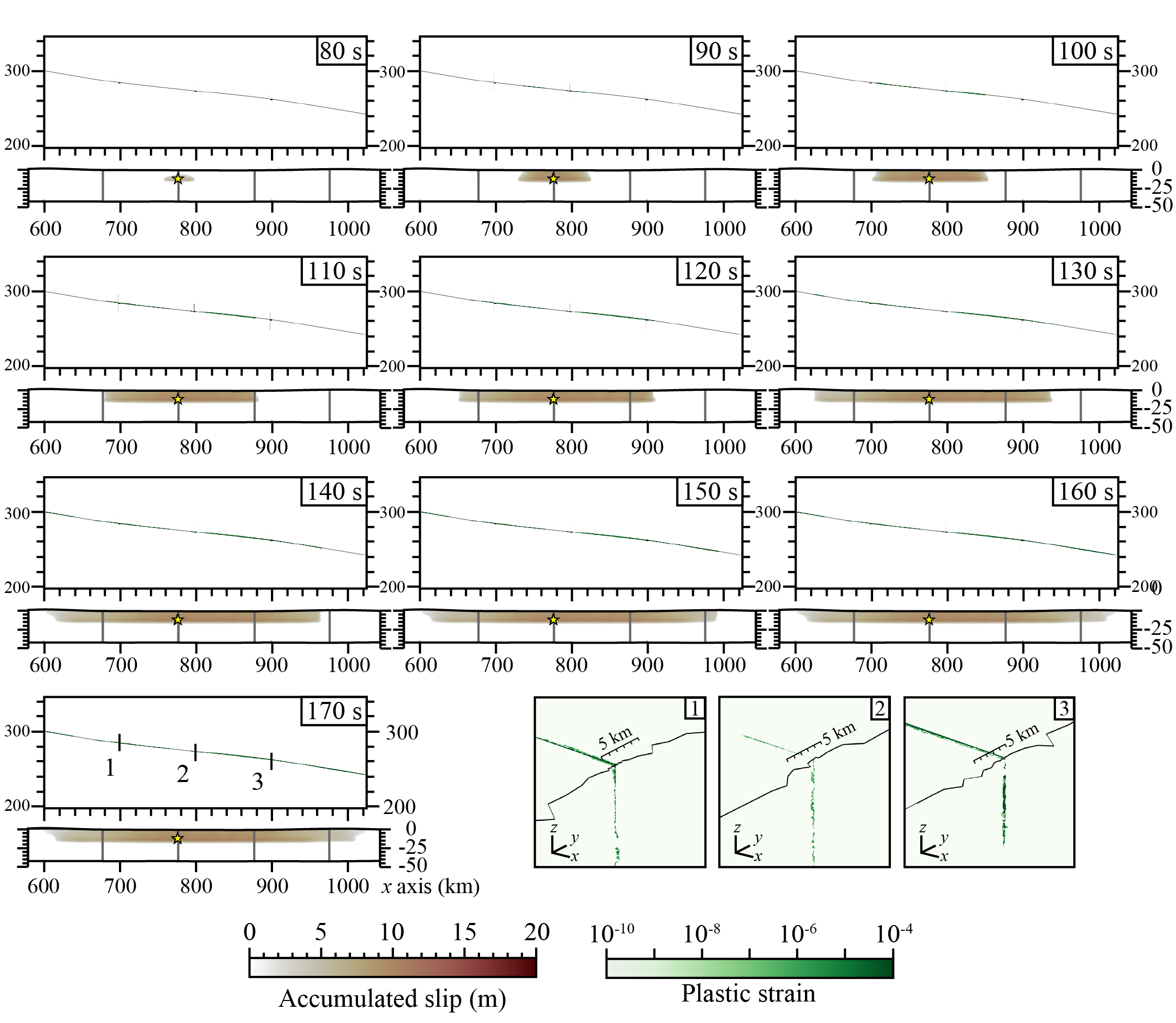}}
\caption{Evolution of the on-fault accumulated slip and off-fault plastic strain for $D_c = 0.6$ m in the model with a quartz upper crust and lower crust.
The trace of the fault (thin black line) is represented in map view with the accumulated plastic strain.
The three zoomed cross-sections (1, 2, 3) are indicated on the map.
Below the map view, a zoom of the fault's surface between $x \approx 600$ km and $x \approx 1000$ km shows the accumulated slip at the given time.
The vertical grey lines are spaced every 100 km starting from the hypocenter.
The length scale between the map and the fault's surface are the same.}
\label{fig:weak-off-plast}
\end{figure}
In this model, we choose a critical slip-weakening distance of $D_c = 0.6$ m, consistent with the value used for the quartz upper crust and anorthite lower crust model (Section \ref{subsubsec:ref-off-fault-damage}). 
The spontaneous rupture initiation occurs slightly before 80 seconds (Figures \ref{fig:weak-off-plast} and \ref{fig:moment-scaling}a), propagating symmetrically away from the hypocenter. 
The propagation of the rupture along the brittle-ductile boundary is ahead of the propagation near the surface until approximately 120 seconds when the rupture at depth reaches the fault's geometrical variations located at $x \approx {600,1000}$ km at 250~km from the hypocenter in both directions (Figure \ref{fig:weak-off-plast}).

The off-fault deformation width around the fault (Figure \ref{fig:weak-off-plast}) is narrower compared to the quartz-anorthite model (Figure \ref{fig:reference-off-plast}), covering only a few hundred meters both at the surface and ar depth. 
The rupture ceases after 170 seconds of simulation time, corresponding to a 90-second long earthquake of magnitude $M_w = 8.02$, with a surface rupture length of 434 km and a rupture surface area of 7336 km$^2$. 
Compared to the rupture generated in the quartz-anorthite model, the full quartz model produces a shorter surface rupture length, a smaller rupture surface area, and less accumulated slip, with an average slip of 6.6 m compared to 9.4 m for the quartz-anorthite model (Table \ref{tab:dr-results}). 
The accumulated slip pattern also shows higher accumulation within the nearest 100 km radius around the hypocenter before progressively decreasing towards the fault's geometrical variations.

Overall, this model generates an earthquake that is qualitatively similar to the earthquake generated in the quartz-anorthite model but with a smaller magnitude.

\subsubsection{Anorthite upper and lower crust (strong rheology)}
\label{subsec:strong-off-fault-damage}
\begin{figure}[htb!]
\centerline{\includegraphics[width=0.99\textwidth]{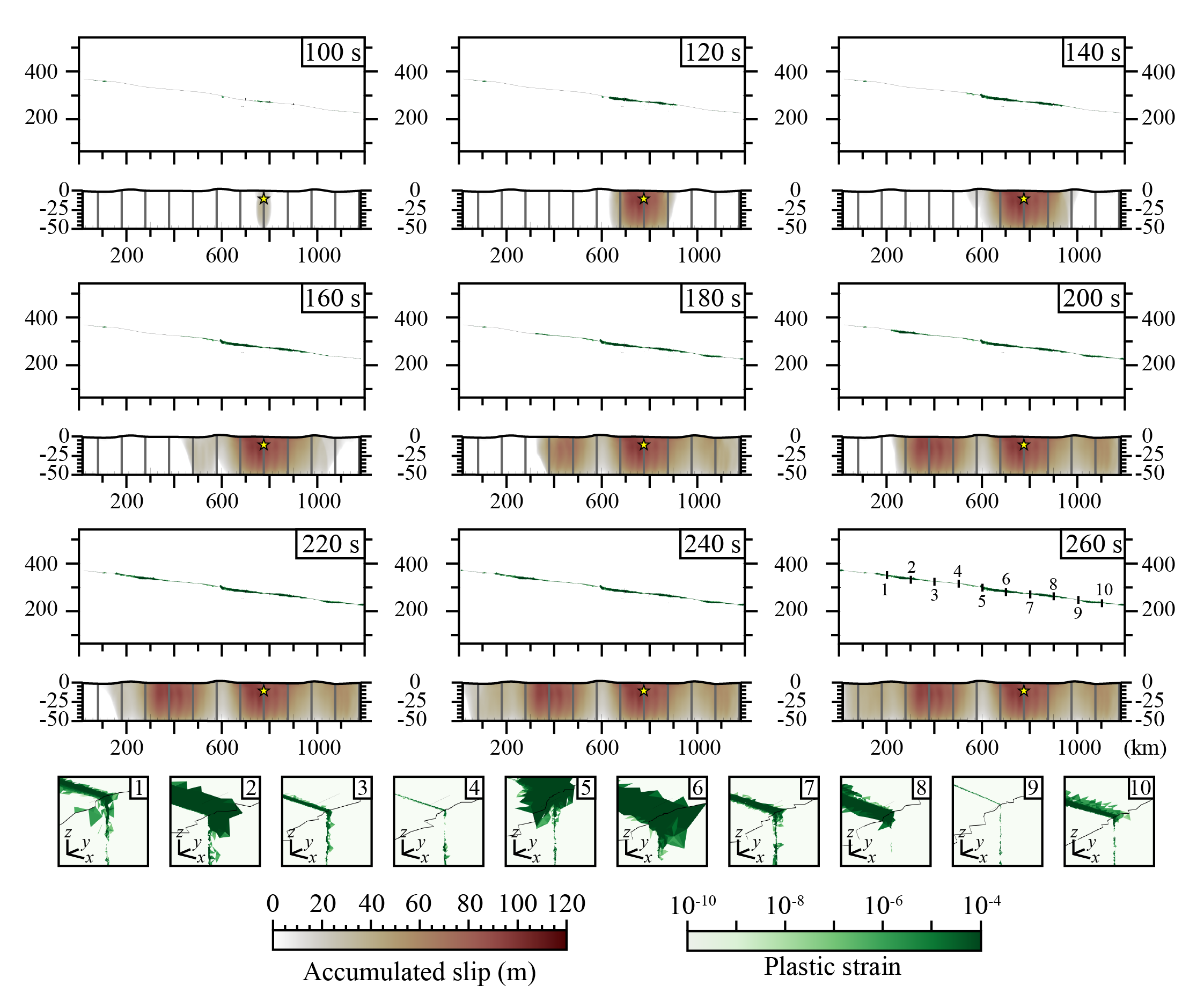}}
\caption{Evolution of the on-fault accumulated slip and off-fault plastic strain for $D_c = 1$ m in the model with an anorthite upper crust and lower crust.
The trace of the fault (thin black line) is represented in map view with the accumulated plastic strain.
The ten zoomed cross-sections (1-10) are indicated on the map.
Below the map view, a zoom of the fault's surface between $x \approx 600$ km and $x \approx 1000$ km shows the accumulated slip at the given time.
The vertical grey lines are spaced every 100 km starting from the hypocenter.
The length scales of the map and the fault's surface are the same, but the fault's surface is vertically exaggerated by a factor of 3.3.}
\label{fig:strong-off-plast}
\end{figure}
This model employs an anorthite long-term rheology for both the upper and lower crust, resulting in significantly higher crustal strength compared to models using a quartz rheology for the upper crust. 
Consequently, there is an absence of the brittle-ductile transition and higher stresses (Figure \ref{fig:strength-ini}) on the fault. 
As a result, the rupture propagates across the entire fault's surface from the surface to a depth of 50 km (Figure \ref{fig:strong-off-plast}).

Upon initiation, dynamic rupture first propagates at depth before extending across the whole fault. 
Initially symmetric, the propagation becomes asymmetric upon reaching the fault's surface geometrical variations located at $x \approx {600, 1000}$. 
The rupture propagates first in the 0 to -25 km depth before progressing in the second half of the fault (-25 to -50 km). 
During passage through the fault's bends, the amplitude of accumulated slip reduces from $\sim 100$ to $\sim 40$ m.
In addition, while rupture passes through the bends, the moment rate shows a significant decrease from $1.8 \times 10^{21}$ Nm.s$^{-1}$ to $0.3 \times 10^{21}$ Nm.s$^{-1}$ (Figure \ref{fig:moment-scaling}c).

In contrast to previous models where rupture is stopped by the fault's geometrical heterogeneity, here, the rupture passes through the pronounced fault bends and continues. 
Passing the fault's bend at $x \approx 600$ km, the amplitude of accumulated slip increases to a magnitude similar to that near the hypocenter ($\sim$ 100 m), and the moment rate increases again to a value of $1.6 \times 10^{21}$ Nm.s$^{-1}$ (Figure \ref{fig:moment-scaling}c). 
This result indicates that the long-term 3D stress field favors slip on fault segments better aligned with the regional plate motion (i.e., long-term boundary conditions). 
At 200 seconds (100 seconds after the rupture initiation), the rupture reaches another fault bend at $x \approx 200$ km, where the magnitude of accumulated slip decreases to approximately 40 m within this region.
By the simulation's end, the entire fault ruptures, with two high-magnitude slip patches corresponding to fault segments better aligned with the plate velocity field.

During rupture, off-fault plastic deformation propagates near the fault within a region reaching up to 10 km in width, illustrated in panels (1) of Figure \ref{fig:strong-off-plast}. 
Similar to the quartz-anorthite model, the off-fault strain exhibits a positive flower-like structure, with a width ranging from a few meters at depth to several kilometers at the surface (Table \ref{tab:dr-results}).

The simulated earthquake has a magnitude of $M_w = 9.34$, generating a surface rupture length of 1165~km, a rupture area of 62,465~km$^2$, and an average slip of 48~m. 
These quantities are exceedingly high compared to natural expectations (Figure \ref{fig:moment-scaling}c, d), attributable to the very strong long-term rheology employed. 
However, this simulation provides valuable insights into rupture behavior when crossing a fault bend and the interplay between stress state, fault geometry, and their influence on rupture dynamics.

\begin{table}
    \centering
    \begin{tabular}{l p{1cm} p{1.5cm} p{1.2cm} p{2cm} p{2cm} p{2cm} p{1cm}}
        \toprule
        Model & OFP & Crust composition & $D_c$ (m) & Average slip (m) & Rupture area (km$^2$) & Surface rupture length (km) & $M_w$ \\
        \midrule
        M1 & No & Q-A & 0.6 & 9.3 & 15,087 & 773 & 8.33\\
        M2 & No & Q-A & 1 & 10.1 & 7,572 & 540 & 8.16\\
        M3 & No & Q-A & 1.5 & 10.2 & 5,619 & 461 & 8.07\\
        M4 & No & Q-A & 1.7 & 10.4 & 5,012 & 386 & 8.04\\
        M5 & No & Q-A & 0.6-1.6 & 9.5 & 9,771 & 600 & 8.21\\
        M6 & No & Q-A & 1.1-2 & 10 & 5,970 & 493 & 8.09\\
        M7 & Yes & Q-A & 0.6 & 9.3 & 9,120 & 550 & 8.19\\ 
        M8 & Yes & Q & 0.6 & 6.6 & 7,336 & 434 & 8.02\\ 
        M9 & Yes & A & 1 & 48.2 & 62,465 & 1,166 & 9.34\\
        M10 & Yes & Q & 0.1 & 8.2 & 21,310 & 1,051 & 8.39\\
        M11 & No & Q & 0.6 & 6.6 & 7,142 & 417 & 8.01\\
        M12 & No & A & 0.6 & 51.8 & 62,944 & 1,165 & 9.37\\
        M13 & No & A & 1 & 48.2 & 62,465 & 1,165 & 9.35\\
        M14 & Yes & Q-A & 1 & 10.4 & 5,611 & 339 & 8.08\\
        \bottomrule
    \end{tabular}
    \caption{Principal characteristics of the rupture for each models. OFP: Off-fault plasticity, Q: quartz, A: anorthite.}
    \label{tab:dr-results}
\end{table}

The coseismic off-fault plastic strain is influenced by the stress state of the long-term model, which depends on the imposed crustal rheology. 
Consequently, in a strong crust model, the stress state more readily reaches plastic yielding during a coseismic rupture in a dynamic model that accounts for off-fault plasticity compared to a weak crust model. 
This results in a wide zone off-fault of high plastic strain values in the strong crust model, unlike for the weak crustal rheology.

\section{Discussion} 
\label{sec:discussion}
\subsection{Relationships between long-term geodynamics and earthquakes dynamics}
\subsubsection{Effect of 3D fault geometry on dynamic rupture propagation}
The reconstructed fault geometry remains simple yet non-planar. 
While the first order orientation of the fault strike is approximately N~100$^\circ$E (west-northwest - east-southeast) over 1000 km, introducing initial weak zones at a $7^{\circ}$ angle with respect to the velocity field introduces a slight obliquity accommodated in long-term deformation by local variations in fault orientation (approximately N~95$^\circ$E). 
These local variations occur every 400 km and span roughly 100 km. 
Despite their small scale, such variations significantly influence the geodynamic characteristics of the system, which in turn impacts the rupture dynamics.

The three long-term geodynamic models, although defined with different rheologies, exhibit a stress regime variation from purely strike-slip to compressional (Figure \ref{fig:long-term-maps}e,i) and a topographic high (Figure \ref{fig:long-term-maps}b,f,j) when transitioning from 400 km long segments to 100 km long geometrical heterogeneity. 
Long-term slip rates also decrease slightly across fault bends (Figure \ref{fig:long-term-maps}d,h,l). 
As our results indicate, this combination affects the 3D stress state of the fault and, consequently, the rupture dynamics during an earthquake.

The dynamic ruptures depicted in Figures \ref{fig:reference-off-plast}, \ref{fig:weak-off-plast}, and \ref{fig:strong-off-plast} all demonstrate that the rupture velocity and accumulated slip decrease near fault bends \cite{Lozos2011,Kase2006,Ma2022}. 
In weaker crusts, where stresses are lower, rupture halts upon reaching fault geometrical heterogeneity (Figures \ref{fig:reference-off-plast} and \ref{fig:weak-off-plast}). 
However, in sufficiently strong crusts, the rupture passes through, behaving similarly on the other side of the heterogeneity.

This behavior highlights how minor variations in the long-term 3D stress field can strongly affect the rupture dynamics, providing a physical mechanism for halting earthquake propagation.

\subsubsection{Relationship between long-term rheology and rupture propagation}
Dynamic rupture models highlight the sensitivity of the rupture propagation to the long-term rheology of the crust. 
The flow laws used to model the continental crust influence both the depth of the brittle-ductile transition and the stress field. 
For instance, the anorthite rheology, gathering feldspar-rich rocks like granulites, gabbros, or diorites that constitute the lower continental crust at first order, undergoes plastic deformation until temperatures of approximately $700^\circ$C (Figure \ref{fig:long-term-setup}). 
This behavior extends the depth of the brittle-ductile transition and thus increases the crustal stress. 
While stresses required for viscous deformation of rocks decrease exponentially with temperature (and thus depth, Eq. \ref{eq:eta-v}), the plastic yield stress criterion increases with depth (Eq. \ref{eq:eta-p}). 
Consequently, the long-term stress field contains this information and significantly influences the thickness of the seismogenic zone and, thus, rupture dynamics.

Models featuring an upper crust composed of quartz, related to granitic-like rocks, exhibit a brittle-ductile transition ranging from 15 km to 20 km depth (Figure \ref{fig:strength-ini}a,b). 
This viscous layer acts as a decoupling layer for long-term crustal mechanical behavior and as a barrier to rupture propagation at earthquake timescales.

Conversely, models with a full anorthite crust composition show no brittle-ductile transition (Figure \ref{fig:strength-ini}c) due to the much greater strength of feldspar-rich rocks compared to quartz-rich rocks. 
The absence of a ductile layer permits the rupture propagation throughout the entire fault thickness. 
Additionally, because feldspar-rich rocks do not readily flow at crustal temperatures, the total stress accumulated along the fault is higher, enabling rupture propagation through geometric variations where fault orientation changes.

Thus, linking long-term geodynamic models with dynamic rupture models establishes mechanical relationships between timescales and assesses the first-order importance of crust composition and long-term rheology in earthquake mechanics. 
This approach allows us to incorporate physics-informed stress states of a lithosphere with a multilayered strength structure into dynamic rupture models. 
Without this method, we would need to explicitly define a stress variation function to represent transitions between brittle and ductile regimes based on compositions that may not always be known at depth. 
This is significant for dynamic rupture models because no single strength profile can represent all types of continental lithosphere \cite{Burov2011}.

\subsubsection{Interpretation of the long-term stress and implication on earthquake dynamics}
The stress field derived from long-term geodynamic models incorporates shear zone geometry, lithosphere rheology, temperature field, volume forces, and topography, resulting directly from momentum and thermal energy conservation given material parameters. 
However, these models compute stresses based on the visco-plastic behavior of the lithosphere over extended periods of time, with typical time steps covering thousands of years. 
Thus, the stresses, re-evaluated at each time step, can be interpreted as the result of loading over this period.

Since our long-term geodynamic models do not account for small time-scale elastic energy dissipation, such as the seismic cycle, the obtained stress values represent the stress state of a system that has not dissipated elastic energy over thousands of years. 
However, in nature, despite the observation of the seismic cycle, large-magnitude earthquakes still occur. 
This suggests that not all accumulated energy dissipates through small-magnitude earthquakes and that long-term visco-plastic lithospheric behaviour and large-scale tectonic plate boundary forces are the first-order drivers responsible for large-magnitude earthquakes. 
As illustrated in Figure \ref{fig:moment-scaling}d, geodynamically informed earthquakes produce high magnitudes, yet for realistic long-term rheologies, they adhere to scaling relationships between $M_w$ and surface rupture length established from natural observations.

In addition, our experiments involve a single fault surface with slight geometrical variations.
Given how these variations strongly impact the stress state and the dynamics of the rupture, accounting for more complicated fault geometries and possible interactions between tectonic structures within a fault network may contribute to dissipating elastic energy more efficiently during the rupture propagation, which would reduce the magnitude of the modelled earthquakes.
Moreover, in the presented models, we imposed the nucleation location due to an absence of a stress state favorable to spontaneous nucleation. 
In addition, the shallow frictional cohesion included with Eq.~\eqref{eq:on-fault-cohesion} prevents the simulation from developing near-surface spontaneous nucleation.
However, as \citeA{VanZelst2019} proposed that long-term geodynamic model pre-stress could predict spontaneous nucleation for dynamic rupture, it is possible that introducing more complexity in the fault network and geometry distributes 3D stresses more consistently to natural cases favoring spontaneous nucleation.

This implies that utilizing long-term geodynamic models to provide information about 3D fault geometry and stress state is valid but should be limited to large-magnitude earthquakes. 
Alternatively, one could use the long-term stress state and fault geometry in a seismic cycle simulation, which performs intermediate-scale elastic energy dissipation, and then transfer the resulting stresses to dynamic rupture simulations.

\subsection{Limitations} 
\label{sub:limitations}

The method proposed in this study is effective for identifying and characterizing first-order localized shear zones. 
However, in 3D long-term geodynamic models, it is not uncommon for major shear zones to be accompanied by diffuse deformation. 
Unfortunately, due to their low strain rate values, highly diffuse deformation or non-localized shear zones cannot be accurately extracted. 
This is because the creation of $\xi$ isovalues surfaces for such cases results in very large volumes, leading to a noisy approximation of the medial axis.

The method's limitations primarily stem from its inability to handle extremely diffuse deformation, as the strain rate values associated with such zones do not allow for the extraction of well-defined and localized shear zones. 
As a result, further developments or alternative approaches may be required to address the characterization of highly diffuse deformation in long-term geodynamic models.

It is important to acknowledge these limitations and consider their implications when applying the proposed method in scenarios where diffuse deformation is significant.

In this study, we link 3D geodynamic models with 3D dynamic rupture simulations that require prescribed nucleation. 
Coseismically, the slip-dependent fault weakening behavior governed by aging law rate-and-state friction is similar to that governed by linear slip-weakening friction \cite<e.g.,>[]{BizzarriCocco2003,Garagash2021,Kaneko2008}.  
Alternative choices for frictional constitutive laws in dynamic rupture simulations, such as rate-and-state friction, may favour less ad-hoc dynamic nucleation. 
3D earthquake cycle simulations use rate-and-state friction to seamlessly integrate spontaneous aseismic nucleation with dynamic rupture \cite{LapustaLiu2009,JiangLapusta2016,Luo2020,MengDuan2023,LiGabriel2024}. 
However, these simulations pose significant methodological and computational challenges, especially when addressing complex interactions of geometry, friction, and structural properties \cite{Jiang2022,Uphoff2023}.

\section{Conclusions} 
\label{sec:conclusion}
This study provides loose coupling between long-term 3D geodynamic models of strike-slip fault evolution --- featuring a single non-planar strike-slip fault --- and dynamic rupture modelling. 
We introduce a new method to extract and reconstruct complex fault surfaces from 3D volumetric shear zones. 
Our key findings are:
\begin{itemize}
    \item Utilizing our method for fault surface reconstruction from volumetric shear zones allows for the evaluation of the long-term slip rate across the faults.
    \item The dynamic rupture models show that for the geodynamic system considered, the geometry of the fault, the rheology of the crust, and the long-term stress-state, a suitable critical slip weakening distance falls within $D_c \in [0.6,1.5]$. 
    \item The long-term rheology significantly influences the stress state and long-term slip rate on the fault, thereby impacting rupture dynamics and plastic strain localization. Crusts with a thicker ductile layer promote a lower stress state that will produce smaller magnitude earthquakes with shorter surface rupture length, smaller rupture surface area, and less accumulated slip. 
    \item The long-term geometry of the fault plays a crucial role in determining the stress regime at locations of geometrical variations, thereby influencing rupture dynamics by favoring slip on fault segments better aligned with the regional plate motion  (i.e., long-term boundary conditions). This behavior highlights how minor variations in the long-term 3D stress field can strongly affect the rupture dynamics, providing a physical mechanism for halting earthquake propagation.
    \item Because feldspar-rich rocks do not readily flow at crustal temperatures, the total stress accumulated along the fault is higher, enabling rupture propagation through geometric variations where fault orientation changes.
    \item Geodynamically informed earthquakes exhibit high magnitudes ($M_w \geq 8$), which we interpret as resulting from a medium where elastic energy is not released by smaller events during the seismic cycle.
\end{itemize}
These findings highlight the interactions between long-term geodynamic processes and short-term earthquake dynamics, shedding light on the importance of considering the long-term mechanics to simulate and understand the dynamic rupture behavior.

\section{Open Research}
The long-term geodynamic models have been performed using pTatin3d, an open-source software publicly available at \url{https://bitbucket.org/dmay/ptatin-total-dev}. 
The git branch used to produce the results is anthony\_jourdon/oblique-nitsche-poissonP. 
Options file to reproduce the models can be found at \url{https://zenodo.org/records/12646159} in ptatin-options\_file.tar.gz with the DOI 10.5281/zenodo.12646158.

The fault surface reconstruction from volumetric shear zone has been performed using an open-source software that we developed and that is publicly available at \url{https://bitbucket.org/jourdon_anthony/ptatin3d-extract-faults-tools}.

Dynamic rupture models were performed using SeisSol, an open-source software publicly available at \url{https://github.com/SeisSol/SeisSol}.
The input files and meshes necessary to reproduce the results can be found at \url{https://zenodo.org/records/12646159} in ptatin3d\_OWL\_Seissol-SimpleFault.tar.xz with the DOI 10.5281/zenodo.12646158.

\acknowledgments
This work has been supported by the European Union’s Horizon 2020 research and innovation programme (TEAR ERC Starting grant no. 852992).
AJ acknowledges additional support by the French government, through the UCAJEDI Investments in the Future project managed by the National Research Agency (ANR) under reference number ANR-15-IDEX-01.
DAM and AAG acknowledge additional support from the National Science Foundation (grant nos. EAR-2121568, OAC-2311208).
AAG acknowledges additional support from Horizon Europe (ChEESE-2P, grant no. 101093038; DT-GEO, grant no. 101058129; and Geo-INQUIRE, grant no. 101058518), from the National Science Foundation (grant nos. EAR-2225286,  OAC-2139536), and the National Aeronautics and Space Administration (grant no. 80NSSC20K0495).

The authors gratefully acknowledge the Gauss Centre for Supercomputing e.V. (https://www.gauss-centre.eu) for providing computing time on the GCS Supercomputer SuperMUC-NG at the Leibniz Supercomputing Centre (https://www.lrz.de) through project pn49ha.
The authors are grateful to the Université Côte d’Azur’s Center for High-Performance Computing (OPAL infrastructure) for providing resources and support through project CT3D.
Dave A. May and Alice-Agnes Gabriel acknowledge financial support from the National Science Foundation (NSF Grant EAR-2121568).

\bibliography{bibliography}

\end{document}